\documentclass[a4paper,fleqn,usenatbib,useAMS]{mnras}

\usepackage{graphicx}	
\usepackage{amsmath}	
\usepackage{amssymb}	
\usepackage{multicol}        
\usepackage{bm}		
\usepackage{pdflscape}	
\usepackage{natbib}
\usepackage{times}
\usepackage[flushleft]{threeparttable}
\usepackage{calc}
\usepackage{rotating}
\usepackage{lscape}
\bibpunct{(}{)}{;}{a}{}{,} 
\usepackage{longtable}
\usepackage{enumitem} 
\usepackage{hyperref}
\usepackage{pifont}
\usepackage{tikz}

\def\oversim#1#2{\lower0.5pt\vbox{\baselineskip0pt \lineskip-0.5pt
     \ialign{$\mathsurround0pt #1\hfil##\hfil$\crcr#2\crcr\sim\crcr}}}

\usepackage[T1]{fontenc}
\usepackage{ae,aecompl}

\usepackage{mathptmx}

\def\apj {{ApJ}}

\def\aap {{A\&A}}
\def\mnras {{MNRAS}}

\def\apjl {{ApJ}}
\def\aj {{ApJ}}

\def\apjs {{ApJs}}

\title[Molecular hydrogen in IPHAS PNe]{Pushing the limits: detecting 
  H$_2$ emission from faint bipolar planetary nebulae in the IPHAS
  sample\thanks{
Based on observations made with the William Herschel Telescope operated
on the island of La Palma by the Isaac Newton Group in the Spanish
Observatorio del Roque de los Muchachos of the Instituto de Astrof\'\i sica
de Canarias.
}}

\author[G.\ Ramos-Larios et al.]{
  G.\ Ramos-Larios$^{1}$\thanks{Contact e-mail:
    \href{mailto:gerardo@astro.iam.udg.mx}{gerardo@astro.iam.udg.mx}}, 
M.A.\ Guerrero$^{2}$\thanks{On sabbatical leave at the Instituto de
Astronom{\'i}a y Meteorolog{\'i}a, Universidad de Guadalajara,
Jalisco, Mexico},
L.\ Sabin$^{3}$ and
E.\ Santamar\'\i a$^{1,4}$
\\
$^{1}$Instituto de Astronom{\'i}a y Meteorolog{\'i}a,
Dpto.\ de F\'\i sica, CUCEI, Universidad de Guadalajara,
Av.\ Vallarta 2602, C.P.\ 44130, Guadalajara, Jalisco, Mexico \\
$^{2}$Instituto de Astrof\'isica de Andaluc\'ia, IAA-CSIC,
Glorieta de la Astronom\'\i a s/n, 18008 Granada, Spain \\
$^{3}$ Instituto de Astronom\'{\i}a,
Universidad Nacional Aut\'onoma de M\'exico,
Apdo.\ Postal 877, C.P.\ 22860, Ensenada, B.C., Mexico \\
$^{4}$ Centro Universitario de Ciencias Exactas e Ingenier\'\i as, CUCEI, 
Universidad de Guadalajara, Av.\ Revoluci\'on 1500, Guadalajara, 
Jalisco, Mexico
}

\date{Last updated 2017 May\ 25; in original form 2017 Jan.\ 21}

\pubyear{2017}

\begin{document}
\label{firstpage}
\pagerange{\pageref{firstpage}--\pageref{lastpage}}
\maketitle

\begin{abstract}
  
We have obtained deep narrowband images in the near-IR H$_2$ $\lambda$2.122
$\mu$m emission line for a sample of 15 faint IPHAS bipolar planetary nebulae
(PNe) to search for molecular material.
H$_2$ emission is found in most of them (14 out of 15), mostly
associated with rings at their equatorial regions and with
their bipolar lobes.  
These detections add to the high occurrence of H$_2$ emission among bipolar
PNe reported in previous works, resulting from the large reservoir of
molecular material in these sources and the suitable excitation conditions
for H$_2$ emission.  
The correlation between detailed bipolar morphology and H$_2$ luminosity
is also confirmed:  bipolar PNe with broad equatorial rings (R-BPNe) have
almost no continuum emission, are H$_2$ brighter and have larger
H$_2$/Br$\gamma$ line ratio than bipolar PNe with pinched equatorial waists
(W-BPNe).
The origin of this dichotomy is unclear.  
The larger size and age of R-BPNe is consistent with shock excitation of
H$_2$, whereas UV pumping is most likely the excitation mechanism in the
smaller and younger W-BPNe, which would explain their lower H$_2$ luminosity.
Although both types of bipolar PNe seem to proceed from the same
progenitor population, this does not imply that R-BPNe descend from
W-BPNe.
Otherwise, we note that some of the H$_2$-weak bipolar PNe harbor
post-common envelope binary systems and symbiotic stars.  
Finally, we suggest that the long-living H$_2$ emission from 
R-BPNe arises from a discrete distribution of compact knots
embedded within the ionized gas at the equatorial region.  

\end{abstract}




\section{Introduction}

Planetary nebulae (PNe) are the descendants of low- and intermediate-mass
stars ($\sim0.8$-8 $M_\odot$) caught in the short transition between the
asymptotic giant branch (AGB) and white dwarf (WD) phases.  
PNe are routinely observed in optical emission lines that reveal
the distribution and physical properties of ionized material, but
these observations are insensitive to the dusty and molecular
material remnant from the AGB phase.
Infrared (IR) and radio observations can be used to detect this component in
order to determine the full extent and outer geometry of PNe, to constrain
the physical conditions in the regions dominated by dust and molecules (e.g.,
clumps and photo-dissociation regions, PDR), and ultimately to investigate
the final episodes of mass-loss during the late AGB phase (see for example
the works by
\citealt{Huggins1996,Young1999,Matsuura2005,Peretto2007,Phillips2011}).

Among the molecular tracers, the $\nu$ = 1$\rightarrow$0 S(1) 2.122
$\mu$m molecular hydrogen (H$_{2}$) emission line (hereafter we will
refer to this line as H$_2$ unless otherwise stated) is one of the
most common in PNe alongside carbon monoxide (CO) emission.
Since the early IR observations of PNe, it has been noted that the H$_{2}$
emission has a higher occurrence rate among bipolar PNe
\citep[see, e.g.,][]{Zuckerman1988}. 
Bipolar PNe have been suggested to descend from the most massive low-
and intermediate-mass stars \citep{Peimbert1983,Corradi1995}.

Although there is a clear association of bipolar PN, type I PN and PN 
with a lower scale height, all pointing to a higher mass progenitor for 
bipolar PNe, it is not clear why the incidence of bipolar PN, $\sim$20\% 
\citep{Parker2006}, is larger than the fraction of  stars that can produce a type I PN 
\citep[$\simeq$3.5 $M_\odot$ main sequence stars, ][]{Karakas2009}.

As such, they exhibit larger and thicker envelopes, with dense equatorial
regions and compact clumps or knots that act as a shield against the UV
radiation from their central stars (CSPNe), preventing the molecules'
dissociation.  
The association between H$_{2}$ emission and the bipolar morphological class
has been thoroughly discussed by \citet{Kastner1996,Kastner1994} who defined
the so-called {\it``Gatley's rule''}, automatically assimilating the presence
of the H$_{2}$ 2.122 $\mu$m emission line to a bipolar morphology.
Although this is generally the case \citep{Guerrero2000}, sensitive
observations reveal H$_{2}$ emission in PNe of morphological types
other than bipolar \citep{Marquez2013,Akras2017}.

It was also soon realized that the brightest H$_2$ emission was found
at the equatorial regions of bipolar PNe with broad equatorial rings 
and a butterfly shape \citep{Webster1988}.  
These regions can be resolved into individual knots and
dense clumps that are embedded within ionized material
\citep[][]{Cox1998,Speck2002,Matsuura2009,Manchado2015,Marquez2015}.
On the other hand, PNe with a pinched-waist and a bow-tie shape have
less prevalent H$_2$ emission which is furthermore associated to a
PDR at their bipolar lobes \citep[e.g., Hb\,12,][]{Dinerstein1988,Hora1996}.  
Following \citet{Marquez2015}, we will refer to the group of
broad-ring bipolar PNe (i.e., \emph{``butterflies''}) as R-BPNe,
and to the group of pinched-waist bipolar PNe (i.e., \emph{bow-ties}
or \emph{hour-glasses}) as W-BPNe.

\begin{table}
\begin{center}
\addtolength{\tabcolsep}{-5pt}
\caption[]{
\label{sample}
Sample of IPHAS PNe observed with WHT LIRIS in the near-IR.
}
\begin{tabular}{|l|l}
\hline
\multicolumn{1}{c}{PN\,G} &
\multicolumn{1}{c}{IPHASX J} \\ 
\hline
G035.4$+$03.4 & \hspace{0.3cm}184336.6$+$034640$^{1}$ \\ 
G038.9$-$01.3 & \hspace{0.3cm}190718.1$+$044056      \\  
G045.7$+$01.4 & \hspace{0.3cm}190954.7$+$120455$^{2}$ \\ 
G045.7$-$03.8 & \hspace{0.3cm}192847.2$+$093436      \\ 
G054.2$-$03.4 & \hspace{0.3cm}194359.5$+$170900      \\ 
G057.9$-$00.7 & \hspace{0.3cm}194226.0$+$214521$^{3}$ \\
G062.7$+$00.0 & \hspace{0.3cm}194940.9$+$261521       \\ 
G062.7$-$00.7 & \hspace{0.3cm}195248.8$+$255359       \\ 
G064.1$-$00.9 & \hspace{0.3cm}195657.6$+$265713       \\ 
G068.0$+$00.0 & \hspace{0.3cm}200224.3$+$304845       \\ 
G081.0$-$03.9 & \hspace{0.3cm}205527.2$+$390359$^{4}$ \\ 
G091.6$-$01.0 & \hspace{0.3cm}212335.3$+$484717       \\ 
G095.8$+$02.6 & \hspace{0.3cm}212608.3$+$542015       \\ 
G101.5$-$00.6 & \hspace{0.3cm}221118.0$+$552841       \\ 
G126.6$+$01.3 & \hspace{0.3cm}012507.9$+$635653       \\ 
\hline
\end{tabular}
\begin{minipage}{9cm}
Other names:$^{1}$ PM\,1-253, $^{2}$Te\,6, $^{3}$Kn 7, $^{4}$TEUTSCH PN\,J2055.4+3903. 
\end{minipage}
\end{center}
\end{table}

\citet{Guerrero2000} investigated the H$_2$ over Br$\gamma$ line ratio in
these two morphological sub-classes and concluded that R-BPNe have higher
H$_2$ to Br$\gamma$ line ratios than W-BPNe.
This is consistent with the H$_2$-dominated and H~{\sc i}-dominated PNe
described by \citet{Hora1999}.  
The origin of this branching among bipolar PNe is 
unclear, but it may be related to the dominant excitation mechanism of 
the H$_2$ molecules in PNe.  
The H$_2$ emission line spectrum can be excited \emph{(i)} by fluorescence 
or radiative pumping through the absorption of the UV radiation from the 
CSPN in a PDR \citep{Black1987,Dinerstein1988,Sternberg1989}, or 
\emph{(ii)} by shocked gas \citep{Burton1992}.  
 
These excitation mechanisms result in notably different  intensities of the H$_2$ 1-0 S(1) line, 
being much brighter for shock-excited H$_2$.
Thus, \citet{Marquez2015} used the very different efficiency of these two
excitation mechanisms to argue that large H$_2$ 1-0 S(1) to Br$\gamma$ 
emission line ratios are indicative of shock excitation.  
They conclude that the H$_2$ brighter R-BPNe are preferentially excited 
by shocks, whereas the H$_2$ weaker W-BPNe are excited by UV fluorescence.

In this paper, we aim to extend the search for H$_2$ emission to PNe
that are fainter than those included in previous works.
If more evolved, these faint bipolar PNe can be used to investigate the
evolution of the molecular content in bipolar PNe.
The UWISH2 survey has detected H$_2$ emission from a number of bipolar PNe
candidates which seems to be highly obscured, as they are not detected in
H$\alpha$ \citep{Froebrich2015,Gledhill2017}.  
Similarly, the INT Photometric H$\alpha$ Survey data 
(IPHAS\footnote{\url{http://www.iphas.org/}}:\citealt{Drew2005,Gonzalez2008,Barentsen2014}) 
has unveiled a number of new faint PNe \citep[e.g.,][]{Viironen2009}.
In particular, the IPHAS catalogue of extended PNe \citep{Sabin2014}
has revealed a set of 159 new true and possible PNe located in the
Galactic Plane in the latitude range $b=|5^\circ|$ among which 45
has been described as bipolar.
This sample can be used to investigate the occurrence of H$_2$ emission
among faint bipolar PNe.
This research has also benefited from a new work on the estimation of 
distances to PNe \citep{Frew2016}.  
Otherwise, by selecting carefully sources whose nebular emission is not 
overimposed on crowded stellar fields, their H$_2$ fluxes can be measured
accurately, contrary to nearby PNe with large angular sizes projecting 
onto multiple field stars.

The paper is organized as follows.
Section \S2 presents the sample and describes the selection criteria.
The details of the new IR and archival optical observations that have been used
in this work are given in section \S3.  
The data reduction
is also explained in this section.
A discussion on the individual H$_2$ morphologies of each PNe and on their
ionized and molecular distributions is presented in \S4, as well as the
different morphological groups which can be drawn from our sample.  
Finally a general discussion, including here the sample from
\citet{Guerrero2000}, and the conclusions are presented in \S5
and \S6, respectively.

\section{Sample}

We selected a set of fifteen PNe from the IPHAS sample of true and possible
PNe \citep{Sabin2014} mostly based on H$\alpha$+[N~{\sc ii}] morphologies
indicative of the presence of bipolar lobes collimated by equatorial rings
or waists.
To maximize their detection, we chose PNe which were optically ``bright''
among IPHAS sources, implying H$\alpha$ dereddened peak fluxes up to
10$^{-14}$ erg~cm$^{-2}$~s$^{-1}$.

Furthermore, to assess the H$_2$ spatial distribution, we selected sources
large enough to be resolved, but not too large so they can fit within the
field of view (FoV) of the instrument used for these observations (see
below).
An additional selection criterium requires minimizing the number of
field stars overimposed on the optical nebular emission.  
Thus our sample includes PNe with optical diameters (major axis)
ranging from $\sim$9\arcsec\ to $\sim$78\arcsec.

The IAU-PNG and IAU-IPHAS designations of the PNe in our sample (the latter
includes their J2000 equatorial coordinates) are listed in columns 1 and 2
of Table\ref{sample}.
Detailed information of these sources can be found in the
HASH database \citep{Parker2016}.
We note that according to the HASH database, G035.4$+$03.4 classification
has become symbiotic star (SS),  and G038.9$-$01.3 and G062.7$-$00.7 are now
flagged as likely PN.
For the especial case of G054.2$-$03.4, a post-common envelope (CE) binary, 
must be treated carefully about our findings.

\section{Observations}

\subsection{Near-IR imaging}

Narrow-band H$_{2}$ 2.122 $\mu$m and $K_c$ continuum near-IR images
were obtained with the Long-slit Intermediate Resolution Infrared
Spectrograph (LIRIS: \citealt{Manchado2003,Acosta2003}) at the 4.2m
William Herschel Telescope (WHT) on Roque de Los Muchachos Observatory
(ORM: La Palma, Spain).
The characteristics of the filters are shown in Table \ref{filters}.
The detector is a 1024$\times$1024 HAWAII array with plate scale of
0\farcs25~pixel$^{-1}$ and an FoV of 4\farcm27$\times$4\farcm27.

\begin{table}
\begin{center}
\caption[]{\label{filters} Filters characteristics for each instrument used in the present study.}
\begin{tabular}{|c|c|c|c|}
\hline
Telescope/Instrument &Filter         &  $\lambda_{c}$ &  $\Delta\lambda$ \\
\hline
 WHT/LIRIS & H$_{2}$   &  2.122 $\mu$m &  0.032 $\mu$m\\
           &  Kc      & 2.270 $\mu$m  & 0.034 $\mu$m\\ 
 \hline
 INT/WFC   & H$\alpha$ & 6568 \AA & 95 \AA \\
\hline   
 \end{tabular}
 \end{center}
\end{table}

\begin{figure*}
  \includegraphics[width=\textwidth]{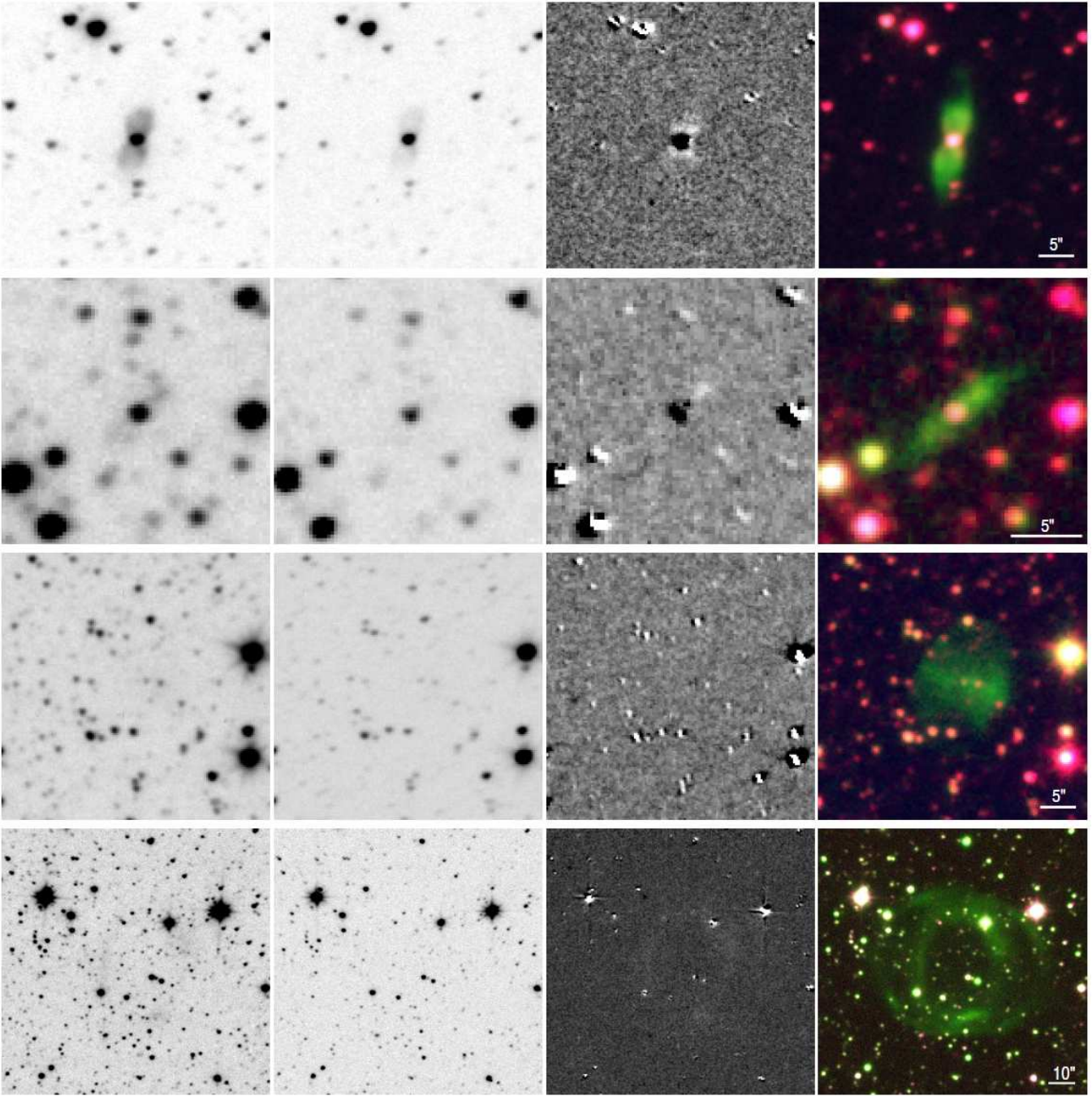}
\caption{
WHT LIRIS near IR images (\emph{from top to bottom}) of PN\,G035.4$+$03.4, PN\,G038.9$-$01.3, PN\,G045.7$+$01.4 and PN\,G045.7$-$03.8 in the H$_{2}$, $K_c$, continuum-subtracted H$_2-K_c$ filters and a colour-composite WHT H$_2$ (red), IPHAS
H$\alpha$+[NII] (green), and WHT $K_c$ (blue) (\emph{from left to right, respectively}). In all images, north is top, east to the left.
The spatial scale is shown in the last colour panel.}
\label{g35}
\end{figure*}

\begin{figure*}
\includegraphics[width=\textwidth]{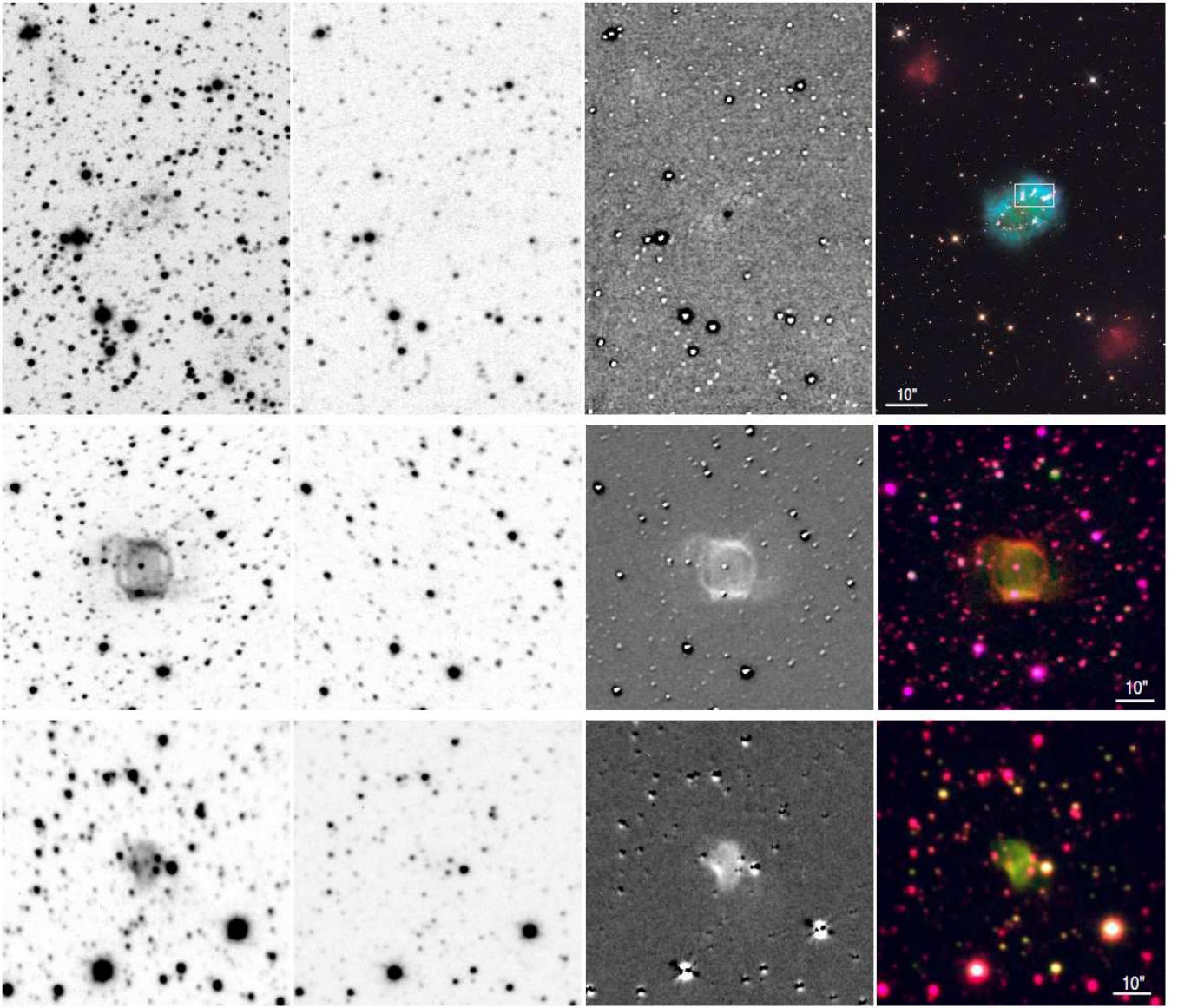}
\caption{Same as Fig. \ref{g35} (\emph{from top to bottom}) for PN\,G054.2$-$03.4, PN\,G057.9$-$00.7 and PN\,G062.7$+$00.0.
In this case, the right panel of PN\,G054.2$-$03.4 show a colour-composite \emph{HST} WFC3 [N~{\sc ii}]
(red), H$\alpha$ (green), and [O~{\sc iii}] (blue) picture.  
The small box in the same panel corresponds to the image section
shown in Figure~\ref{neckzoom}.
\emph{HST} image courtesy of the Hubble Heritage Team (STScI/AURA).  
\label{neck}
}
\end{figure*}

\begin{figure*}
\includegraphics[width=\textwidth]{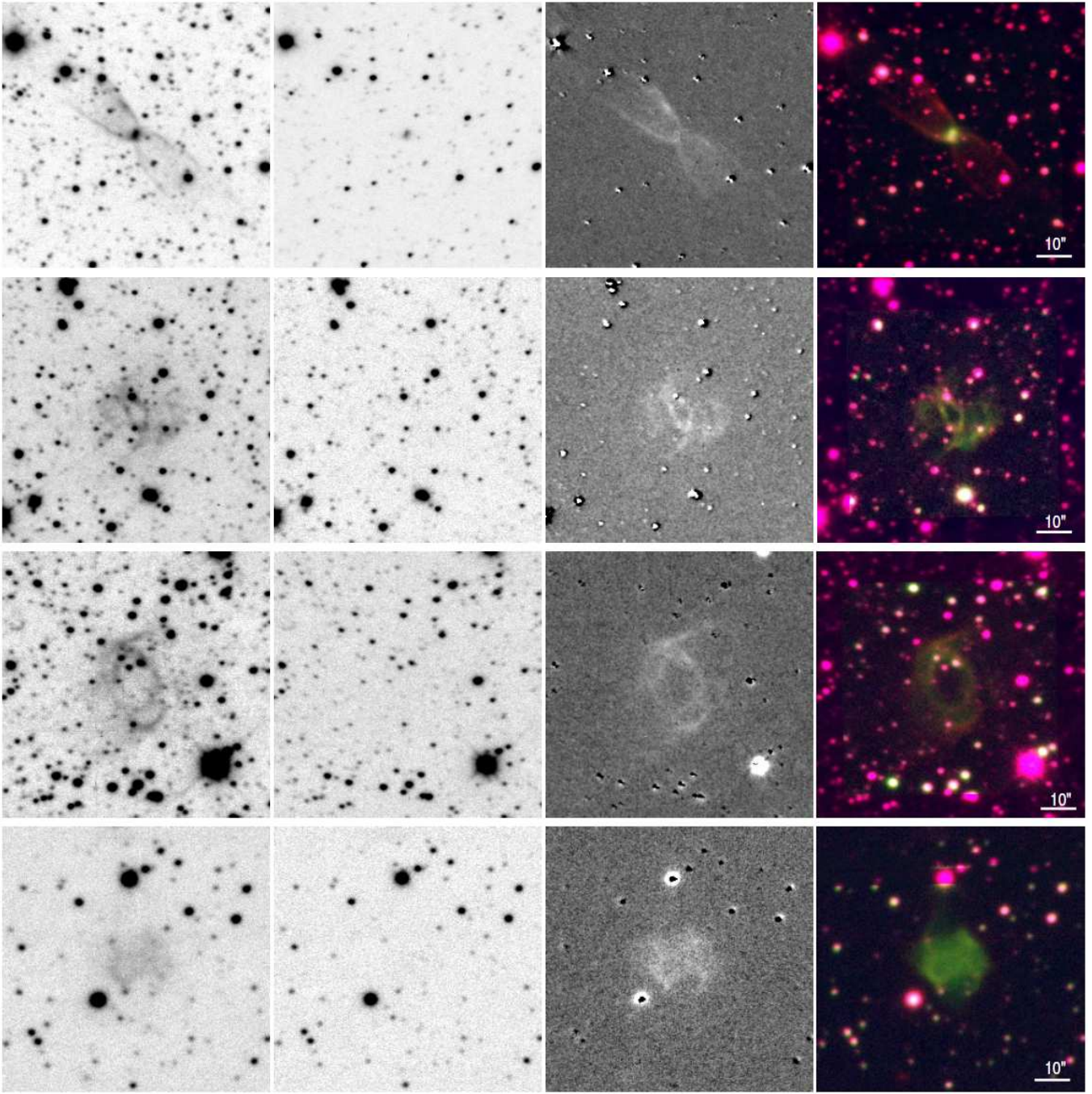}
\caption{Same as Fig. \ref{g35} (\emph{from top to bottom}) for PN\,G062.7$-$00.7, PN\,G064.1$-$00.9, PN\,G068.0$+$00.0 and PN\,G081.0$-$03.9.}
\label{g62_2}
\end{figure*}

\begin{figure*}
\includegraphics[width=\textwidth]{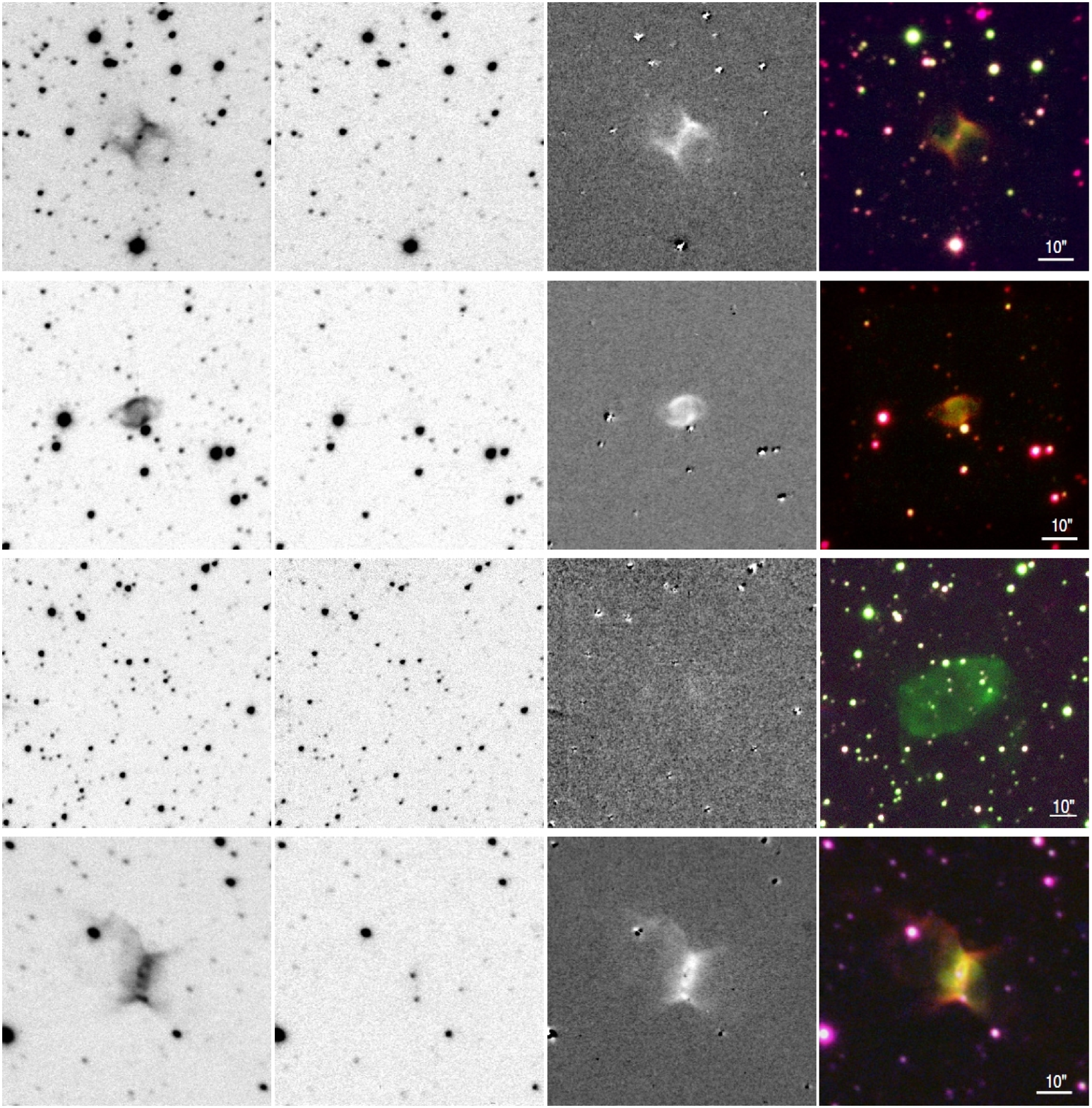}
\caption{Same as Fig. \ref{g35} (\emph{from top to bottom}) for PN\,G091.6$-$01.0, PN\,G095.8$+$02.6, PN\,G101.5$-$00.6 and PN\,G126.6$+$01.3.}
\label{g91}
\end{figure*}

Most observations were carried out on 2014 July 13.
One object, IPHASX\,J194359.5+170900 (a.k.a., the Necklace nebula,
\citealt{Sabin2008,Corradi2011,Miszalski2013}) has been previously
observed on the same telescope on 2007 September 19.
We acquired 16$\times$60s exposures in H$_{2}$ and 8$\times$60s exposures
in $K_c$ for each object during the most recent run, and 150$\times$20s
exposures during the 2007 run.  
The IR observing procedure which includes images jittering has been
described in details by \citet{Ramos2012}\footnote{
See also the WHT/LIRIS website
  at:\\~\url{http://www.ing.iac.es/Astronomy/instruments/liris/imaging.html}}.
The data reduction was performed using the package {\sc lirisdr} (LIRIS Data
Reduction), which is an {\sc iraf}\footnote{
{\sc iraf} (Image Reduction and Analysis Facility) is distributed by the
National Optical Astronomy Observatories, which are operated by the
Association of Universities for Research in Astronomy, Inc., under
cooperative agreement with the National Science Foundation.
}-based pipeline dedicated to the automatic
reduction of near-IR data.
The processing steps include bad pixel mapping, cross-talk correction,
flat-fielding, sky subtraction, removal of reset anomaly effect, field
distortion correction, and co-addition to form the final image.
The procedure is repeated for each H$_{2}$ and $K_c$ data-sets separately.  
The latter, which corresponds to the continuum emission, is
subsequently scaled and subtracted from its H$_{2}$ image
counterpart.
The new near-IR images are presented in Figures~\ref{g35}, \ref{neck}, \ref{g62_2} and \ref{g91}
arranged by their Galactic longitude.

\begin{figure*}
\includegraphics[width=2.0\columnwidth]{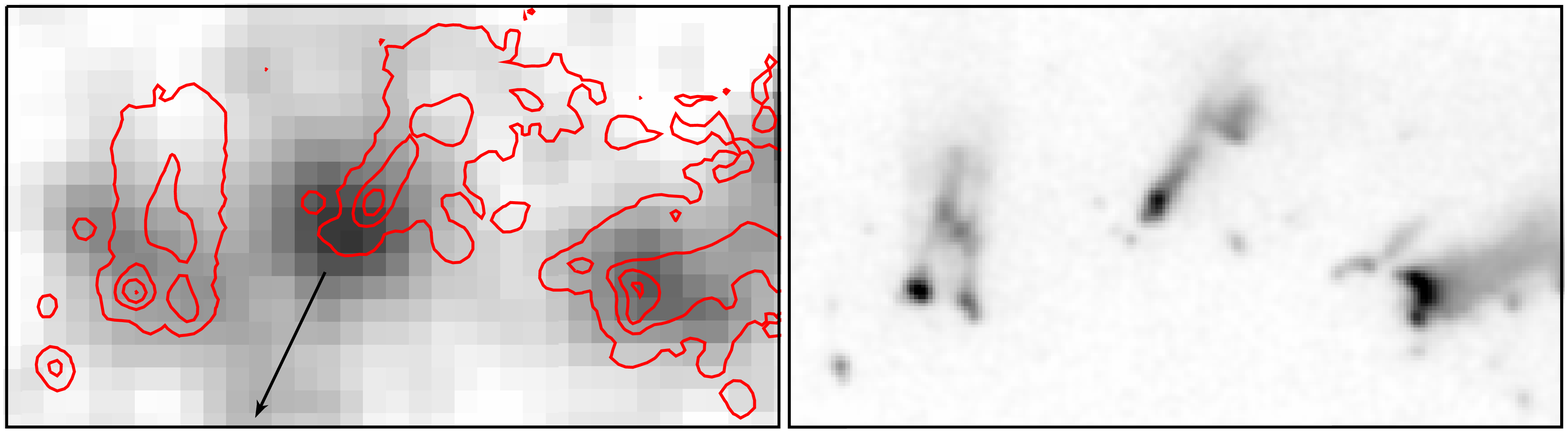}
\caption{Zoom of the North-western knots in the ring of the Necklace
  Nebula in H$_2$ (left) and [N~{\sc ii}] (right).
  [N~{\sc ii}] contours are overplotted on the H$_2$ image.
  The arrow in the H$_2$ image points to the central star.
  The FoV is 8\farcs5$\times$5\farcs0.  
  \label{neckzoom}}
\end{figure*}

It is worth emphasizing the high throughput and efficiency of the
LIRIS instrument at the WHT in terms of detection of molecular
hydrogen at low surface brightness levels.
M\,2-48 and NGC\,6778 were reported as H$_{2}$-free by \citet{Kastner1996}
and \citet{Webster1988}, respectively, but recent WHT LIRIS observations
actually detected H$_2$ emission \citep{Marquez2013}.
This is critical for this project, as we initially deal with even
fainter optical PNe than M\,2-48 and NGC\,6778.

\subsection{Optical imaging}

The optical images presented in this article were
retrieved from the IPHAS images data bank\footnote{
\url{http://www.iphas.org/images/}}.  
These images were originally obtained with the Wide Field Camera (WFC)
mounted on the 2.5m Isaac Newton Telescope (INT), also located at the
ORM.
The camera has a FoV of 34$\times$34 arcmin$^{2}$ and a plate scale of
0\farcs33~pixel$^{-1}$.
The properties of the H$\alpha$ filter used during the survey are
given in Table~\ref{filters}.  
We note that the H$\alpha$ filter also contains the [N~{\sc ii}]
$\lambda$6583 \AA\ emission line \citep{Drew2005}.
We will therefore refer to these as H$\alpha$+[N~{\sc ii}] images.
Exposure times were always 120 s.

These images were combined with the near-IR images to create the
colour-composite pictures presented in the \emph{right}
panels of Figures~\ref{g35}, \ref{neck}, \ref{g62_2} and \ref{g91}.

\section{Basic morphology in the near-IR H$_2$ line}

All the sources in our sample, but PN\,G045.7$+$01.4 (a.k.a.\ Te\,6),  are
detected in H$_2$.  
Te\,6 is a barrel-like PN with broad \emph{ansae} which can be interpreted
as bipolar lobes.  
The nebula has a diameter of 17\farcs3$\times$13\farcs8 in the H$\alpha$
line (Fig.~\ref{g35}).  
The optical emission is mostly localized in a broad, diffuse equatorial
region, whereas the polar protrusions (particularly the Southern one)
are much fainter.

According to the morphology and intensity of the H$_2$ emission, the
sources in our sample with detected H$_2$ emission can be organized
into two different groups.  
PNe with equatorial pinched waist (i.e., W-BPNe) have typically faint H$_2$
emission (e.g., PN\,G035.4+03.4, Fig.~\ref{g35}), whereas those with a well
developed equatorial ring (i.e., R-BPNe) tend to have bright H$_2$ emission
(e.g., PN\,G126.6+01.3, Fig.~\ref{g91}).
These two groups are described into more detail below.

\subsection{W-BPNe: PNe with pinched equatorial waists}

This group includes PN\,G035.4+03.4, G038.9$-$01.3
(Figure~\ref{g35}), and G062.7$-$00.7 (Figure~\ref{g62_2}).
The optical morphology of these PNe is characterized by narrow waists in
between a pair of bipolar lobes (PN\,G035.4+03.4), a highly collimated
bipolar outflow (PN\,G038.9$-$01.3), and a pair of hourglass-shaped bipolar
lobes (PN\,G062.7$-$00.7).

In the near-IR continuum $K_c$ filter, they show mostly compact, bright
emission in the innermost regions, whereas the images in the H$_2$ filter
show additional extended emission along the collimated structures.
(Fig.~\ref{g35}, and \ref{g62_2} \emph{left and middle-left}).  
The continuum subtracted images (Fig.~\ref{g35}, and
\ref{g62_2} \emph{middle-right}) reveal the lack of 2.122 $\mu$m
H$_{2}$ emission in the innermost regions, being the molecular
emission mostly restricted to the collimated structures.  
Otherwise, the details of the diffuse H$_2$ emission differs among these three
sources.

\noindent
$\bullet$ PN\,G035.4$+$03.4 (a.k.a.\ PM\,1-253) \\
The 2.122$\mu$m H$_{2}$ emission traces the onset of the bipolar ejections,
much smaller in size than the $\sim$9\farcs6 emission detected in the
H$\alpha$+[N~{\sc ii}] IPHAS image (Fig.~\ref{g35}).  
The eastern side of the equatorial plane seems to present very faint 
H$_{2}$ emission, but we reckon it may be an artifact caused by a
{bf 
defective 
}
subtraction of the bright continuum emission.  
The distribution of the H$_{2}$ region is somehow similar to that
described for M\,2-9 \citep{Phillips1985,Kastner1996}, although
the detailed morphology is not that close.
The H$_2$ emission in M\,2-9 follows the walls of the bipolar lobes in
their full extent, what has been referred to as a ``thin H$_{2}$ skin''
\citep{Hora1994,Smith2005}, whereas in the ``IPHAS counterpart''
PN\,G035.4+03.4 it is concentrated at the base of the lobes.

\noindent
$\bullet$ PN\,G038.9$-$01.3 \\
The molecular hydrogen emission is mostly detected along the collimated
outflow, with the north-west component of the outflow  being brighter than
the south-east one (Fig.~\ref{g35}).
The asymmetry of the H$_2$ emission could be interpreted as the result of
asymmetric mass loss, but also produced by an orientation effect causing
higher extinction to the south-east region.
In this sense, the H$\alpha$+[N~{\sc ii}] IPHAS image also reveals
brighter emission of the north-west region.  
The nebular morphology of PN\,G038.9$-$01.3 is rather similar to that of
M\,1-91 \citep{Guerrero2000}.  
The latter has been classified as an SS \citep[e.g.][]{Schmeja2001},
whereas PN\,G038.9$-$01.3 is flagged as a likely PN in the HASH
database \citep{Parker2016}.  

\noindent
$\bullet$ PN\,G062.7$-$00.7 \\
The H$_{2}$ emission follows the ionized emission seen in the light of
H$\alpha$+[N~{\sc ii}] (Fig.~\ref{g62_2}).  
The H$_{2}$ data indicate a total size of $\sim$61\farcs5.
The H$_2$ emission from the central region is located in the
outermost region, enclosing the pinched waist.

In many ways, 
the spatial distribution of the H$_2$ emission from this
IPHAS PN that encloses the optical emission from the bipolar outflows is
similar to that shown in the WHT continuum-subtracted H$_{2}$ 2.122 $\mu$m
image of the well-known bipolar PN NGC\,6537 \citep{Marquez2013}.
The HASH database casts doubts on the PN nature of this source
and flags it as a likely PN \citep{Parker2016}.

\subsection{R-BPNe: PNe with broad equatorial rings}

This group includes the remaining PNe in our sample, namely
PN\,G045.7$-$03.8, G045.7$-$03.8, G057.9$-$00.7, G062.7$+$00.0,
    G064.1$-$00.9, G068.0$+$00.0, G081.0$-$03.9, G091.6$-$01.0,
    G095.8$+$02.6, G101.5$-$00.6, and G126.6$+$01.3.
These PNe are characterized by the presence of large equatorial rings
or broad structures that are at the origin of bipolar lobes or other
axisymmetrical structures such as ansae or protrusions.

In the near-IR, these sources show basically no emission in the $K_c$
continuum images.
The diffuse H$_2$ emission is spatially coincident with the optically
brighter equatorial structure and, in a few cases, with the fainter
bipolar components.  
The intensity of the H$_2$ emission echoes that seen in the
H$\alpha$+[N~{\sc ii}] IPHAS images, with brighter emission
in the equatorial regions.
We describe briefly the H$_2$ morphology of some individual sources.

\noindent
$\bullet$ PN\,045.7$-$03.8 \\
The optical images show two interleaved circular rings
$\sim$58\farcs1$\times$68\farcs2 in size with brighter
emission in the central regions where they overlap.
Weak H$_2$ emission is detected in the central regions, spatially
coincident with the brightest patches in the H$\alpha$+[N~{\sc ii}]
IPHAS image (Fig.~\ref{g35}).

\noindent
$\bullet$ PN\,G054.2$-$03.4 (a.k.a.\ the Necklace Nebula) \\
The optical images of the Necklace Nebula show a knotty ring and a pair of high
velocity collimated outflow/jets.
These have been presumably launched from an accretion disk before the
common envelope phase of the close binary system formed by its
central star and a carbon dwarf companion \citep{Corradi2011,Miszalski2013}.  
The 2.122 $\mu$m H$_2$ emission arises exactly at the location of
the ionized knots, but also at the polar caps (Figure~\ref{neck}).  
The H$_{2}$ surface brightness in this PN is extremely low and its
detection required an integration time about three times longer than
that for the other objects in our sample.

The detailed image of the Northwest quadrant of the equatorial ring
of the Necklace Nebula shown in Figure~\ref{neckzoom} is revealing.
The H$_2$ emission comes from a discrete distribution of compact knots and
filaments associated with the [N~{\sc ii}] knots and their tails, rather
than from a classical PDR.  
This is very similar to the distribution of the H$_2$-emitting knots
embedded in the equatorial regions of NGC\,650-51 \citep{Marquez2015}
and NGC\,2346 \citep{Manchado2015}, but also in the low-ionization
features (FLIERs) of NGC\,7662 \citep{Akras2017}.
It is worth emphasizing the varying spatial distribution of the molecular and
low-ionization material in these knots; sometimes the H$_2$ emission in the
knot faces the central star and the [N~{\sc ii}] emission lies beyond (for
instance, the central knot in Fig.~\ref{neckzoom}), but in some other cases
the H$_2$ emission seems to peak further away from the central star than the
[N~{\sc ii}] emission (as in the two other knots in Fig.~\ref{neckzoom}).
This may reveal projection effects, suggesting that the H$_2$
and [N~{\sc ii}] emissions are not coplanar.  
Certainly, an H$_2$ image of spatial resolution similar to that of
the \emph{HST} WFC [N~{\sc ii}] image is required to investigate
the details of the varying spatial distribution of molecular and
low-ionisation material in this nebula.

\noindent
$\bullet$ PN\,G057.9$-$00.7 (a.k.a.\ Kn\,7) \\
The optical images reveal a bipolar PN showing a bright central ring
partially filled with ionized material and faint bipolar lobes.  
The H$_2$ emission follows the spatial location of the H$\alpha$+[N~{\sc ii}]
emission, being particularly bright in the ring and relatively fainter in its
interior (Figure~\ref{neck}).
The bipolar lobes are also detected in the H$_2$ image.

\noindent
$\bullet$ PN\,G062.7$+$00.0 \\
The optical image shows a butterfly nebula with size
$\sim$18\farcs5$\times$11\farcs0 and a bright wide waist.  
The H$_{2}$ emission largely covers the whole ionized region, with stronger
emission in the waist (Fig.\ref{neck}).

\noindent
$\bullet$ PN\,G064.1$-$00.9 and PN\,G068.0$+$00.0 \\
Bipolar PNe with an equatorial ring and two diffuse bipolar outflows.
There is an excellent spatial match between the H$_2$ and optical
H$\alpha$+[N~{\sc ii}] morphologies, with the equatorial ring being
the brightest structure (Fig.\ref{g62_2}).

\noindent
$\bullet$ PN\,G081.0$-$03.9 \\
Bipolar PN with an open barrel-like structure from whose tips protrude
fainter bipolar extensions.  
The H$_{2}$ emission is rather diffuse and located at the barrel-like feature,
i.e., the brightest region of the optical H$\alpha$+[N~{\sc ii}] emission
(Fig.\ref{g62_2}).

\noindent
$\bullet$ PN\,G091.6$-$01.0 and PN\,G126.6$+$01.3 (a.k.a.\ ``Principe de
Asturias'' Nebula) \\
Bipolar PNe with a classical \emph{butterfly} morphology viewed side-on
(Figs.~\ref{g91}).
The central rings are thus seen mostly as bars or very elongated ellipses.
Both nebulae exhibit stars located very close to the centre of the 
nebular equatorial regions.  
Molecular hydrogen is detected along the central bars, enveloping the tips
of the equatorial rings, as well as at the outer walls of the bipolar lobes.

\noindent
$\bullet$ PN\,G095.8$+$02.6 \\
Bipolar PN with a remarkable point-symmetric morphology.  
The spatial distribution of the molecular hydrogen confirms this
point-symmetric morphology (Fig.\ref{g91}).  
Molecular emission is detected in the main nebular shell, and faintly
at the \emph{ansae}.

\noindent
$\bullet$ PN\,101.5$-$00.6 \\
The optical image displays an equatorial structure and two fainter
bipolar lobes.  
The emission in the H$\alpha$+[N~{\sc ii}] IPHAS image is suggestive
of a broad equatorial belt seen from the side, so that the bipolar
lobes lie on the plane of the sky. 
The H$_{2}$ emission is extremely poor and only a patchy feature
is seen towards the brighter north-western edge of the optical
emission (Fig.\ref{g91}).

\section{Discussion}

\subsection{General properties of the H$_2$ emission in the IPHAS sample of PNe}

This study confirms the prevalence of H$_2$ emission among bipolar PNe:
only one source (PN\,G045.7$+$01.4) out of 15 bipolar PNe in this sample
shows no trace of H$_2$ emission in its continuum-subtracted H$_2$ image
(Fig.~\ref{g35}).  
 
Narrow-band near-IR imaging of a combined sample of 141 PNe 
\citep[][this work]{Kastner1996,Guerrero2000,Marquez2015,Akras2017} 
support this prevalence: 
$\sim$75\% (66 out of 89) of bipolar PNe are detected in H$_2$, 
whereas only $\sim$25\% (13 out of 52) of non-bipolar PNe are 
detected in H$_2$.

The H$_2$ emission originates from the equatorial regions and
bipolar lobes of the PNe in the IPHAS sample.  
Whenever there is a broad ring-like structure (i.e., in the 11
R-BPNe with H$_2$ emission), the H$_2$ emission is brighter in
these regions than in the bipolar lobes.
This is the case even for the faint source PN\,G101.5$-$00.6
whose bipolar lobes are most likely missing detection.  
In the remaining three sources that show a pinched-waist (the
W-BPNe sources PN\,G035.4$+$03.4, G038.9$-$01.3, and
G062.7$-$00.7), the H$_2$ emission is brighter in the bipolar
lobes, whereas the central regions are dominated by continuum
emission.
Continuum emission in R-BPNe is basically negligible, 
pointing to the absence of dust thermal emission.

When the spatial distributions of the emission from molecular and ionized
material can be compared, it can be seen that the innermost regions close
to the central star are dominated by ionized material.
Molecular material, as traced by the H$_2$ emission, is confined
to the outer edges of the ionized regions \citep{Beckwith1978}.

The preferential detection of H$_2$ emission associated with the equatorial
regions of bipolar PNe and its distribution at the outer edges of these
confirm that equatorial density enhancements provide ideal conditions to
shelter the H$_{2}$ molecule from UV radiation and avoid its dissociation.
However, the detailed comparison between the molecular H$_2$
and low-ionization [N~{\sc ii}] emissions for the North-west
quadrant of the equatorial ring of the Necklace Nebula
(Fig.~\ref{neckzoom}) suggests that the H$_2$ emission arises
from knots and filaments which are spatially coincident with
the [N~{\sc ii}] knots.
When deep H$_2$ and optical observations of the highest spatial
resolution are obtained, they unveil that the molecular material
in the equatorial regions of bipolar PNe is embedded within
ionized gas \citep[e.g.,][]{Manchado2015}.

\subsection{Prevalence of H$_2$ or H~I emission.}

The prevalence of the emission from the H$_2$ 1-0 S(1) line over 
Br$\gamma$ is commonly used to classify sources into molecular- or
ionized-dominated \citep[e.g.,][]{Hora1999}.  
This classification is useful to investigate the evolutionary stage of the
nebula \citep[e.g.,][]{Aleman2011,Likkel2006} or possible different
evolutionary paths \citep{Guerrero2000}.

As noted in Section~\S1, this same ratio has been associated with
different bipolar morphologies: R-BPNe show brighter H$_2$ than
Br$\gamma$, whereas W-BPNe have brighter Br$\gamma$ than H$_2$
\citep{Guerrero2000,Marquez2015}.

In this context, it is worthwhile to investigate the relative intensity of
the H$_2$ 1-0 S(1) and Br$\gamma$ emission lines in the IPHAS sample and
compare them to those presented by \citet{Guerrero2000}.

The H$_2$ 1-0 S(1) fluxes of the PNe in our sample can be derived from
the WHT LIRIS images.  
First, we have used an aperture encompassing the whole nebular extent of
each PN to derive the total count rate.  
This has been also estimated for a number of stars in the field of
view using {\sc iraf} DIGIPHOT routines.
The count rates from these stars are then compared to their
2MASS $K_s$ magnitudes to derive a photometric zero point
magnitude.  
Then, taken into account the filter transmission curve to derive the
filter equivalent width (EW), the count rates of the nebular emission
in the H$_2$ 1-0 S(1) line have been converted into observed fluxes.
For bright sources, the flux uncertainty is dominated by the Poisson
statistics and it may range from 5\% to 20\%.
For faint and extended sources, the flux uncertainty can amount up to 50\%.

Similarly, the IPHAS images can be used to determine the Br$\gamma$ fluxes
of these sources.  
The images are used to obtain the count rates in the INT
H$\alpha$+[N~{\sc ii}] filter.
Optical spectroscopic observations obtained for these sources in the IPHAS
framework \citep{Sabin2014} are used in conjunction with the filter
transmission curve to assess and subtract the contribution of the [N~{\sc ii}]
$\lambda\lambda$6548,6584 lines to the emission detected in the IPHAS
H$\alpha$+[N~{\sc ii}] image.
The net H$\alpha$ count rate is then converted into observed flux\footnote{
No spectrum in the spectral range covering the H$\alpha$ and [N~{\sc ii}]
emission lines is available for PN\,G035.4$+$03.4.  
The H$\alpha$ flux for this nebula has not been corrected from the
[N~{\sc ii}] contribution, and thus it provides an upper limit of
its true H$\alpha$ flux.
}
using the zero point magnitude $z_p$ provided by \citet{Barentsen2014}.  
This flux is corrected from extinction using the H$\beta$ logarithmic
extinction coefficient, $c$(H$\beta$).
This is derived from the observed H$\alpha$ to H$\beta$ flux ratio measured in
the IPHAS optical spectra compared to the theoretical ratio expected for Case~A
at 10,000 K \citep{Osterbrock2006}.  
For those sources with no available optical spectra covering the H$\alpha$
and H$\beta$ emission lines (namely, PN\,G035.4$+$03.4, G045.7$-$03.8,
G054.2$-$03.4, G062.7$+$00.0, and G062.7$-$00.7), $c$(H$\beta$) was computed
from their H~{\sc i} column density provided by \citet{Dickey1990} and 
\citet{Kalberla2005} assuming the gas-to-dust ratio from \citet{Bohlin1978}.
The intrinsic Br$\gamma$ flux is then derived using the theoretical
Br$\gamma$ to H$\alpha$ flux ratio of 0.00990 which can be derived
using \citet{Osterbrock2006} theoretical line ratios for case~A at
10,000 K.
To allow a fair comparison between the H$_2$ 1-0 S(1) and
Br$\gamma$ fluxes, the flux from the former line is also
dereddened using the measured $c$(H$\beta$).

The intrinsic H$_2$ and Br$\gamma$ fluxes and the [N~{\sc ii}]
$\lambda$6584 to H$\alpha$ and H$_2$ to Br$\gamma$ ratios are
collected in Table~3.
The detailed bipolar morphological sub-types (R-BPNe and W-BPNe)
are also included in this Table.  
For completeness, the sample of bipolar PNe in \citet{Guerrero2000} 
has been added to this table.
The H$_2$ flux of NGC\,6881 listed by \citet{Guerrero2000} 
is considered to be a lower limit in view of the deeper H$_2$
images presented by \citet{Ramos2008}.

\begin{figure*}
{\includegraphics[bb=41 170 540 650,height=7.6cm]{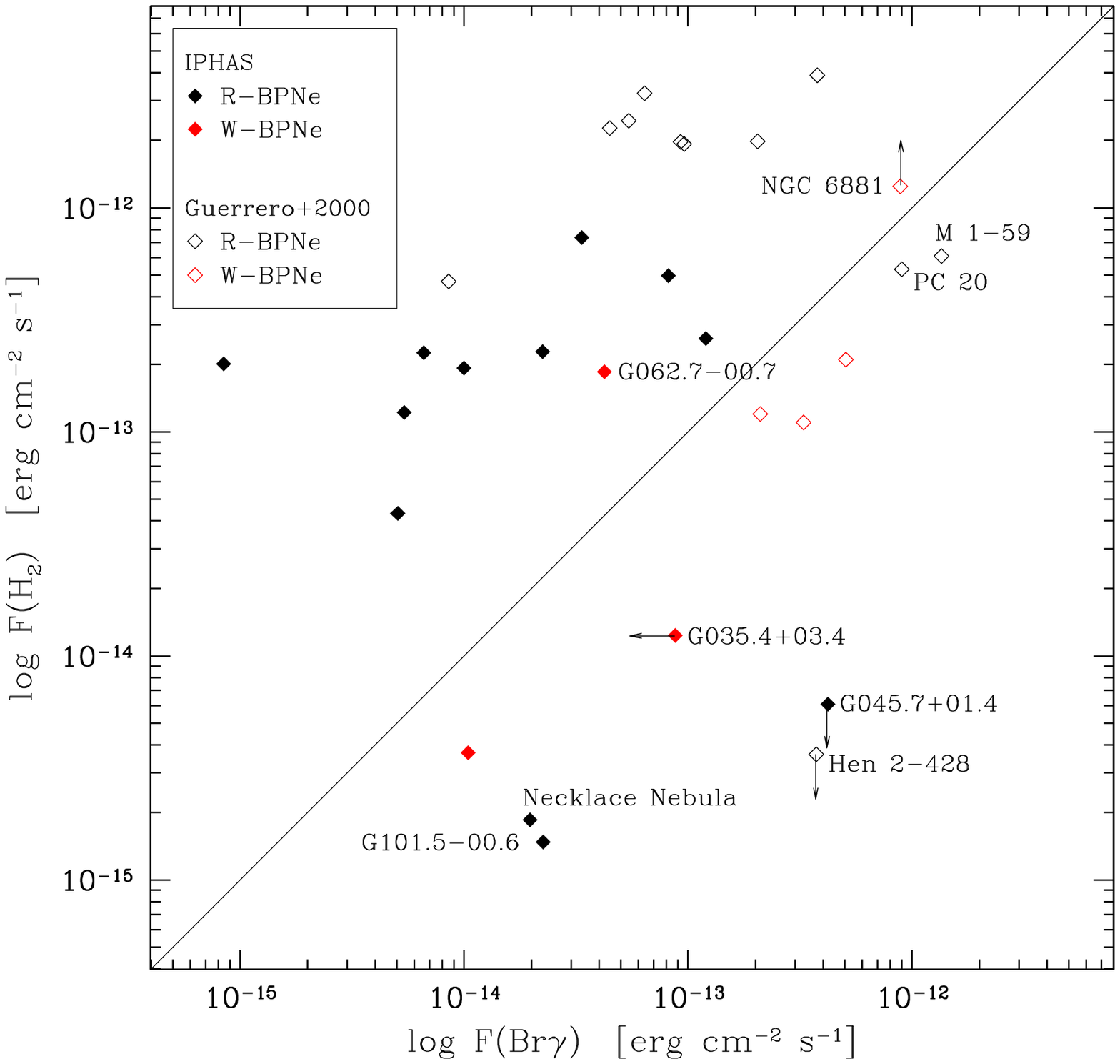}} 
\hspace*{0.5\columnsep}%
{\includegraphics[bb=41 170 540 650,height=7.6cm]{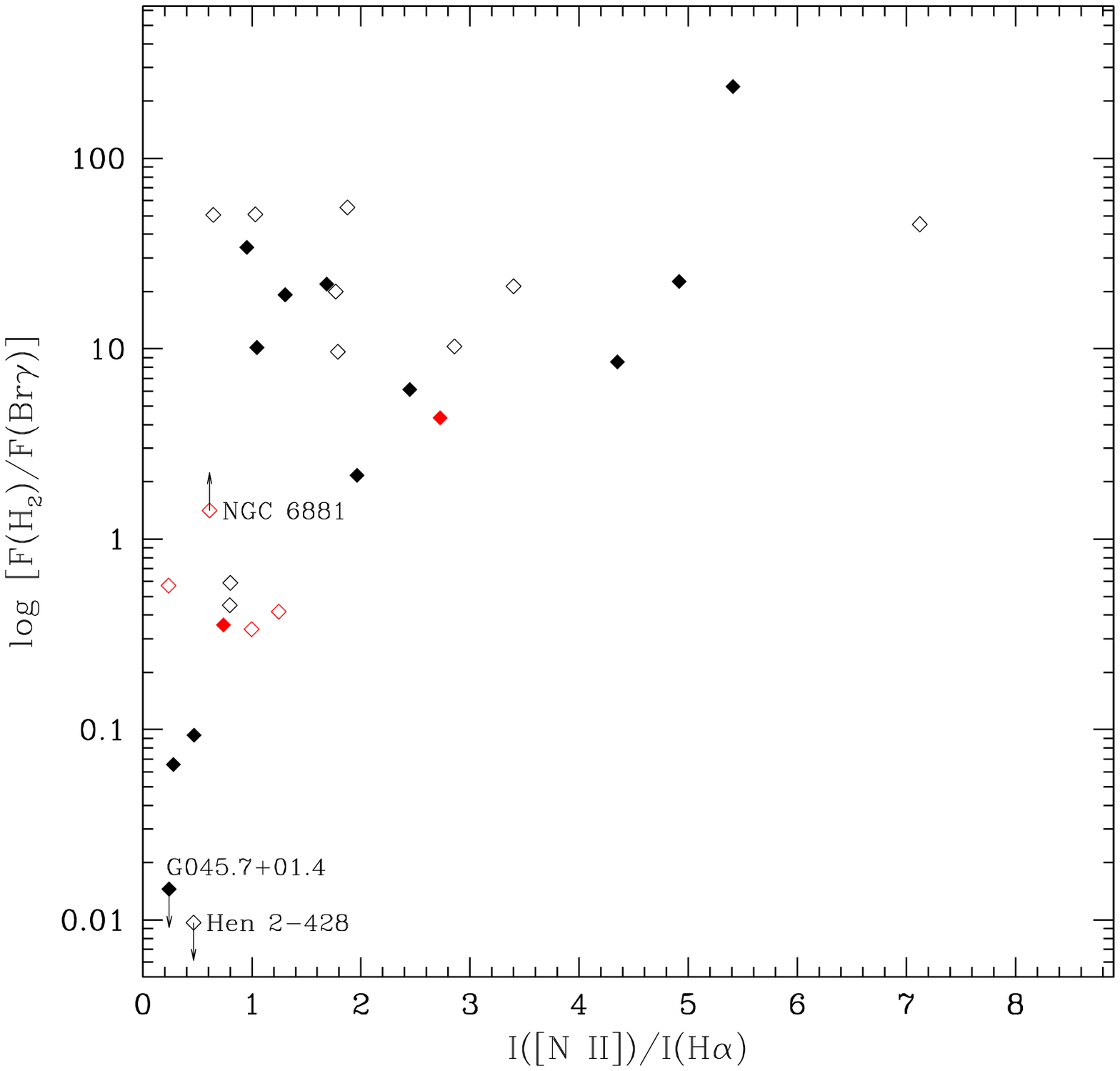}} 
\caption{\label{plot1}
Comparison of the H$_2$ and Br$\gamma$ unabsorbed fluxes \emph{(left)} and
the H$_2$/Br$\gamma$ vs.\ [N~{\sc ii}]/H$\alpha$ line ratios \emph{(right})
of the IPHAS PNe in this work and \citet{Guerrero2000}'s sample.  
Different symbols are used for the different data samples and morphologies
as described in the left panel.
Singular objects are labeled.  
}
\end{figure*}

The trends are shown in Figures~\ref{plot1}, \ref{gal}, and \ref{plot2}.  
The new IPHAS PNe extend towards regions of lower H$_2$ and
Br$\gamma$ fluxes (Figure~\ref{plot1}-{\it left}).  
In agreement with previous results, R-BPNe have generally larger H$_2$ than
Br$\gamma$ fluxes, whereas W-BPNe are typically fainter in H$_2$ than in
Br$\gamma$.
All R-BPNe are detected in H$_2$ with the only exception of Hen\,2-428 and
PN\,G045.7$+$01.4. 
Most R-BPNe have H$_2$/Br$\gamma$ ratios above unity, with the exceptions
of PC\,20, M\,1-59, the Necklace Nebula, and PN\,G101.5$-$00.6.  
Most W-BPNe have H$_2$/Br$\gamma$ ratios below unity, with the only
exceptions of NGC\,6881 and PN\,G062.7$-$00.7.
These exceptions are labeled in Figure~\ref{plot1}-{\it left}.

There is also a notable correlation between the near-IR H$_2$/Br$\gamma$ and
optical [N~{\sc ii}]/H$\alpha$ line ratios (Figure~\ref{plot1}-{\it right});
sources with larger H$_2$/Br$\gamma$ line ratios tend to show also larger
[N~{\sc ii}]/H$\alpha$ line ratios.
Furthermore, R-BPNe do not only have the largest H$_2$/Br$\gamma$
ratios, but they also show the largest [N~{\sc ii}]/H$\alpha$ line
ratios.
Most R-BPNe have H$_2$/Br$\gamma$ line ratios $\geq$10, thus
implying shock excitation \citep{Marquez2015}.  
Similarly, the low H$_2$/Br$\gamma$ line ratio of most W-BPNe implies
UV excitation.

\begin{table*}
\begin{center}
\begin{minipage}{192mm}   
\caption{H$_2$ fluxes for the IPHAS and Guerrero+2000 samples}
\begin{threeparttable}
\begin{tabular}{@{}lccrrrrccccl@{}}
\hline
Common name & PN G & Radius &   d\tnote{a}~~~ & F(Br$\gamma$) & F(H$_2$) & H$_2$/Br$\gamma$ & [N~{\sc ii}]/H$\alpha$ & BPN type & $z$ & L(Br$\gamma$) & L(H$_2$) \\
            &      &  (pc)  & (kpc) & 
\multicolumn{2}{c}{($\times$ 10$^{15}$ erg~cm$^{-2}$~s$^{-1}$)} &
 &  &  &  (pc) &
\multicolumn{2}{c}{($\times$ 10$^{-30}$ erg~s$^{-1}$)}  \\
\hline
\hline

PM\,1-253  & G035.4$+$03.4 & 0.29 &  6.9~ & $<$88~~~ &	    12.3  &   $>$0.14~~  &     $\cdots$ &	W	&	407.32	&	499.57	 &	70.06	\\
$\cdots$   & G038.9$-$01.3 & 0.35 & 17.1~ &  10.4    &       3.7  &	 0.35~~  &	0.74	&	W	&	387.61	&	366.60	 &	130.50	\\
Te\,6      & G045.7$+$01.4 & 0.26 &  3.4~ & 420~~~   &    $<$6.1  &   $<$0.015   &	0.24	&	R	&	83.39	&	590.82	 &	8.58	\\
$\cdots$   & G045.7$-$03.8 & 1.82 &  6.4~ &   5.1    &	    43.2  &	 8.5~~~~ &	4.35	&	R	&	426.54	&	25.26	 &	215.85	\\
Necklace   & G054.2$-$03.4 & 0.72 &  6.9~ &  20~~~   &	     1.9  &	 0.093	 &	0.47	&	R	&	409.57	&	113.46	 &	10.67	\\
Kn\,7      & G057.9$-$00.7 & 0.53 &  6.8~ &  34~~~   &	   730~~~ &	21.9~~~~ &	1.69	&	R	&	82.94	&	186.46	 &	4088.80	\\
$\cdots$   & G062.7$+$00.0 & 0.28 &  6.1~ & 120~~~   &	   260~~~ &	 2.2~~~~ &	1.96	&	R	&	6.42	&	539.58	 &	1172.70	\\
$\cdots$   & G062.7$-$00.7 & 0.45 &  6.9~ &  42~~~   &	   185~~~ &	 4.4~~~~ &	2.73	&	W	&	84.57	&	245.34	 &	1071.03	\\
$\cdots$   & G064.1$-$00.9 & 1.04 &  9.8~ &   5.4    &	   120~~~ &	22.6~~~~ &	4.92	&	R	&	154.03	&	62.59	 &	1419.02	\\
$\cdots$   & G068.0$+$00.0 & 1.88 & 15.3~ &   0.8    &	   200~~~ &    239.1~~~~ &	5.41	&	R	&	9.89	&	23.84	 &	5704.47	\\
$\cdots$   & G081.0$-$03.9 & 0.84 &  8.6~ &  10~~~   &	   190~~~ &	19.3~~~~ &	1.30	&	R	&	581.67	&	88.32	 &	1701.08	\\
$\cdots$   & G091.6$-$01.0 & 0.74 & 11.8~ &   6.6    &	   230~~~ &	34.0~~~~ &	0.96	&	R	&	205.12	&	110.26	 &	3758.30	\\
$\cdots$   & G095.8$+$02.6 & 0.44 &  9.8~ &  22~~~   &	   230~~~ &	10.1~~~~ &	1.05	&	R	&	444.29	&	259.48	 &	2634.17	\\
$\cdots$   & G101.5$-$00.6 & 0.93 &  5.3~ &  23~~~   &	     1.8  &	 0.065	 &	0.28	&	R	&	54.98	&	74.89	 &	4.91	\\
Pr\'\i ncipe de Asturias                                                                                                                         
           & G126.6$+$01.3 & 0.43 &  5.1~ &  82~~~   &	   500~~~ &	 6.1~~~~ &	2.45	&	R	&	115.37	&	255.43	 &	1556.61	\\
\hline
M\,1-57    & G022.1$-$02.4 & 0.20 &  4.0~ & 510~~~   &	   210~~~ &	 0.41~~  &	1.25	&	W	&	165.33	&	950.56	 &	395.28	\\
M\,1-59    & G023.9$-$02.3 & 0.16 &  2.9~ &1350~~~   &	   610~~~ &	 0.45~~  &	0.80	&	R	&	115.98	&	1,361.64 &	615.26	\\
M\,2-46    & G024.8$-$02.7 & 0.24 &  4.2~ & 330~~~   &	   110~~~ &	 0.33~~  &	1.00	&	W	&	196.58	&	687.67	 &	231.33	\\
PC\,20     & G031.7$+$01.7 & 0.14 &  3.8~ & 900~~~   &	   530~~~ &	 0.58~~  &	0.80	&	R	&	113.21	&	1,582.68 &	932.02	\\
M\,4-14	   & G043.0$-$03.0 & 0.45 &  4.6~ &  96~~~   &    1920~~~ &	20.0~~~~ &	1.77	&	R	&	242.89	&	249.97	 &	4994.10	\\
Hen\,2-428 & G049.4$+$02.4 & 0.38 &  2.6~ & 370~~~   &    $<$3.6  &   $<$0.010  &	0.47	&	R	&	110.38	&	312.99	 &	3.05	\\
K\,3-34    & G059.0$+$04.6 & 1.22 &  6.9~ &   8.5    &     470~~~ &	55.2~~~~ &	1.88	&	R	&	549.84	&	48.31	 &	2667.92	\\
M\,1-91    & G061.3$+$03.6 & 0.28 &  4.5~ & 210~~~   &	   120~~~ &	 0.57~~	 &	0.24	&	W       &	285.19	&	523.17	 &	298.96	\\
M\,1-75    & G068.0$-$00.0 & 0.39 &  2.6~ & 380~~~   &    3890~~~ &	10.3~~~~ &	2.86	&	R	&	1.86	&	304.23	 &	3139.09	\\
K\,3-58    & G069.6$-$03.9 & 0.46 &  6.7~ &  45~~~   &	   270~~~ &	50.9~~~~ &	1.03	&	R	&	453.59	&	239.55   &	12192	\\
NGC\,6881  & G074.5$+$02.1 & 0.17 &  3.3~ & 880~~~   & $>$1250~~~ &   $>$1.4~~~~ &	0.61	&	W	&	122.24	&	1,188.06 &	1679.95	\\
M\,4-17    & G079.6$+$05.8 & 0.66 &  4.1~ &  64~~~   &    3240~~~ &	50.6~~~~ &	0.65	&	R	&	415.75	&	130.81	 &	6622.27	\\
K\,4-55    & G084.2$+$01.1 & 0.65 &  4.5~ &  54~~~   &    2450~~~ &	45.1~~~~ &	7.13	&	R	&	86.54	&	133.01	 &	6012.67	\\
M\,2-52    & G103.7$+$00.4 & 0.38 &  3.6~ & 200~~~   &	  1980~~~ &	 9.7~~~~ &	1.79	&	R	&	25.06	&	317.51	 &	3081.68	\\
Bv\,5-1    & G119.3$+$00.3 & 0.46 &  4.6~ &  92~~~   &	  1970~~~ &	21.3~~~~ &	3.40	&	R	&	24.19	&	237.91	 &	5077.89	\\
 \hline
 \hline
 \end{tabular} 
 \begin{tablenotes}
            \item[a] All distances computed according to \citet{Frew2016}
        \end{tablenotes}
     \end{threeparttable}
\end{minipage}
\end{center}
\end{table*}

\subsection{The H$_2$ Luminosity Distribution and Physical Properties of BPNe}

\begin{figure*}
  \includegraphics[height=10cm]{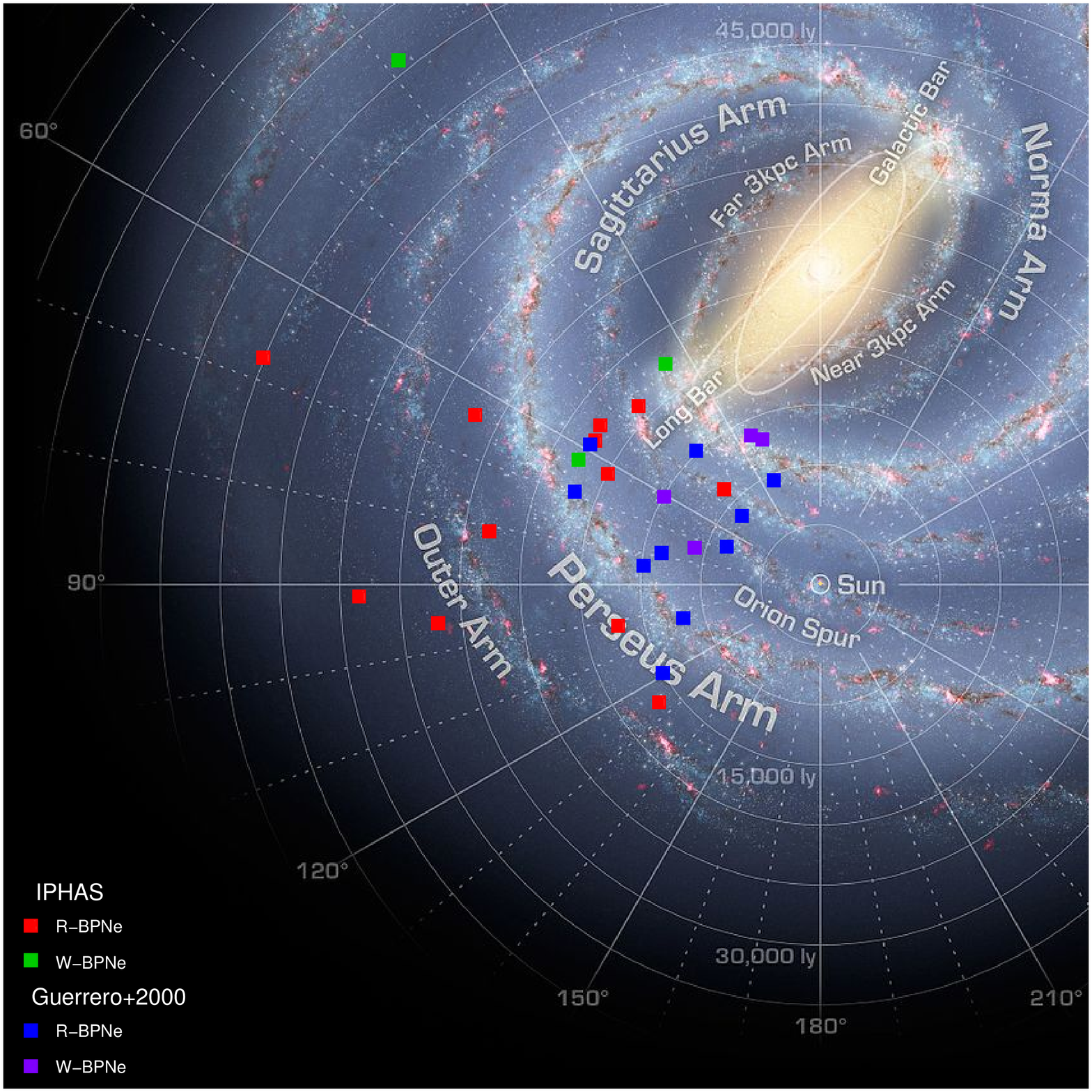}
  \caption{\label{gal}
Distribution of the PNe in our sample in the Galaxy.
The PNe have been located on an artistic view of the Milky Way
(Image credit: NASA/JPL-Caltech/R.\ Hurt SSC/Caltech).  
    }
\end{figure*}

The H$\alpha$ intrinsic flux of each source has been combined with
its size according to the prescriptions of \citet{Frew2016} to derive
its distance.
The distribution of the sources in the Galaxy is shown in Figure~\ref{gal}.  
Whereas the sources sampled by \citet{Guerrero2000} are generally closer than
5 kpc (4.2$\pm$1.2 kpc in average), the IPHAS sources are mostly distributed
above 5 kpc from the Sun (8.4$\pm$3.8 kpc in average), with some of them as
far as 15 kpc.  
This results confirms that the IPHAS sample of bipolar PNe presented
in this work probes sources at further distances in the Galaxy.  
Meanwhile, the faintest H$\alpha$ bipolar PNe sampled by UWISH2
\citep{Froebrich2015} correspond to very extincted sources or
even to sources that have not been ionized yet \citep{Gledhill2017}.

\begin{figure*}
{\includegraphics[bb=41 170 540 650,height=7.6cm]{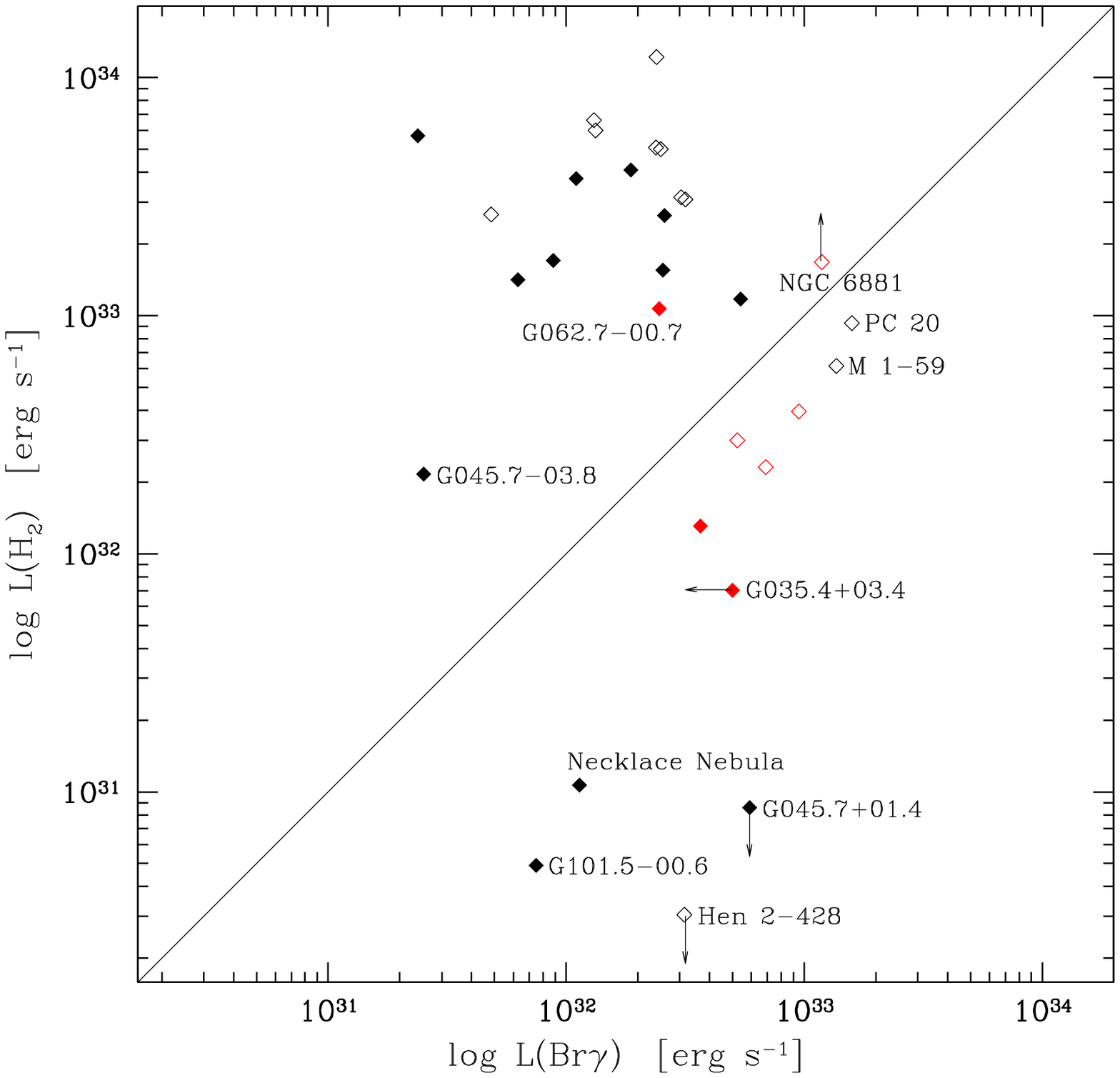}} 
\hspace*{0.5\columnsep}%
{\includegraphics[bb=41 170 540 650,height=7.6cm]{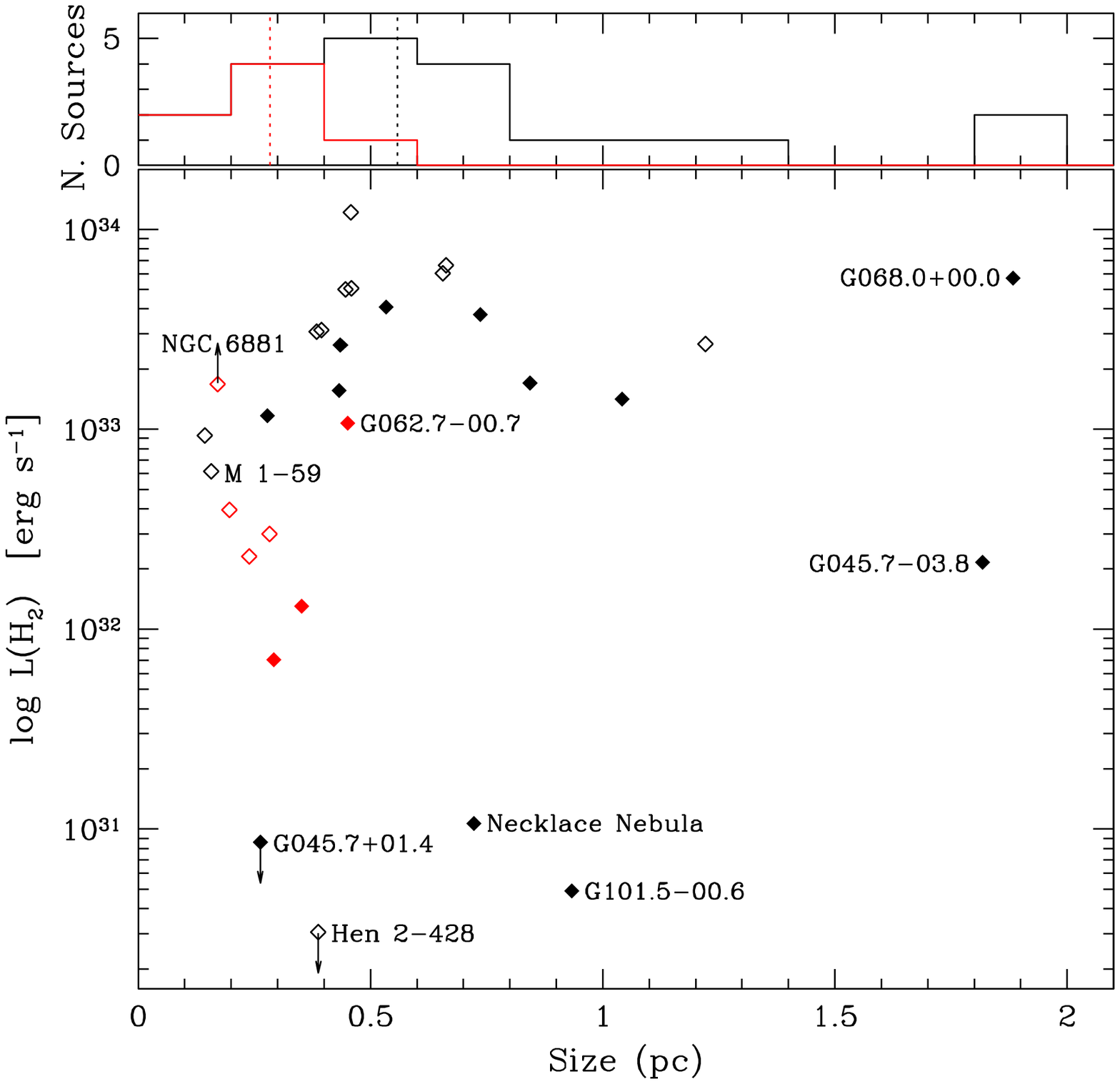}} 
\caption{\label{plot2}
Comparison of the H$_2$ and Br$\gamma$ intrinsic luminosities \emph{(left)}
and distribution of nebular sizes with H$_2$ luminosities \emph{(right}). 
Different symbols are used for the different data samples and morphologies 
as described in the left panel of Figure~\ref{plot1}.
Singular objects are labeled.  
}
\end{figure*}

The distance is then used to derive the luminosity in the H$_2$ 
and Br$\gamma$ emission lines of these BPNe, as well as their
Galactic height and linear size.
The comparison between the H$_2$ and Br$\gamma$ luminosities
(Figure~\ref{plot2}-\emph{left}) confirms that both samples
of bipolar PNe are rather similar, with the IPHAS sample being
generally only a bit fainter than the bipolar PNe sampled by
\citet{Guerrero2000}.  
The low H$\alpha$ surface brightness of the bipolar PNe in the IPHAS sample
does not imply these are more evolved sources, but rather more distant and
extincted.

W-BPNe have typically larger Br$\gamma$ luminosities than R-BPNe,
whereas the opposite is true for H$_2$.  
There are two outstandingly faint H$_2$ R-BPNe in this plot, namely the
Necklace Nebula and PN~G101.5$-$00.6.  
The former has been claimed to be a post-AGB/post-PN ejecta nebula
\citep{Miszalski2013}.

As for the correlation between PN size and H$_2$ luminosity
(Figure~\ref{plot2}-\emph{right}), W-BPNe (0.28$\pm$0.10 pc)
are notably smaller than R-BPNe (0.56$\pm$0.29 pc).  
The latter show a notable flat distribution of H$_2$ luminosity
with size, i.e., the nebular size does not affect the H$_2$
luminosity.
If nebular size can be correlated with nebular age (assuming similar
expansion velocities), then the above result implies that there is
very little evolution of the total H$_2$ luminosity with age and
expansion for R-BPNe.  
As H$_2$ emission is still detectable (and bright) in very extended and
presumably old PNe, it indicates that molecular hydrogen is likely to
survive during a significant fraction of the life of an R-BPNe, only
declining in the final stages of their nebular evolution.

The ``longevity'' of the H$_2$ emission in R-BPNe is most likely linked
with the true nature of molecular material in evolved bipolar PNe.  
The H$_2$ emission does not come from a global PDR, but from a number of
compact knots distributed into equatorial regions pervaded by ionized
material.  
Molecular material survives in these knots and the H$_2$ molecule
is mostly excited by shocks.
This is in agreement with the high-resolution images of the molecular
material in NGC\,2346 \citep{Manchado2015}.  
Similar structures are identified in the classical \emph{butterfly}
PN NGC\,650-51, where the diffuse H$_{2}$ emission is mostly tracing
a ring which is composed of a set of knots and filamentary structures
\citep{Marquez2015}.

\subsection{Bipolar PNe: two of a kind?}

The results reported here confirm the dichotomy between
bipolar PNe with broad rings and those with pinched
waists first reported by \citet{Webster1988}.
\citet{Guerrero2000} questioned the origin of these two different branches of
bipolar PNe, but it could not conclude whether W-BPNe and R-BPNe are different
in nature or are the same kind of sources, but found at different evolutionary
stages.

\begin{figure}
{\includegraphics[bb=71 180 520 700,width=\columnwidth]{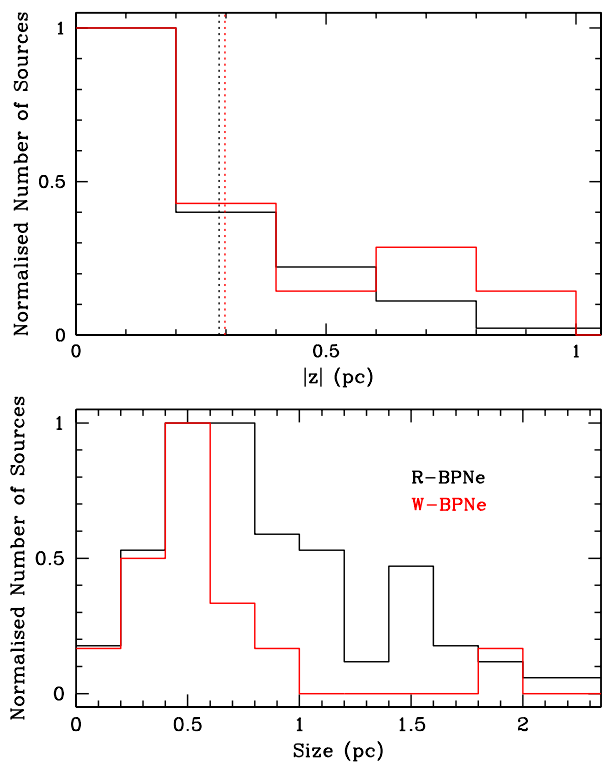}} 
\caption{\label{bpn}
Distribution of the Galactic height (\emph{top}) and size (\emph{bottom}) 
of R-BPNe (black) and W-BPNe (red).
The dotted vertical lines in the top panel correspond to the mean
Galactic heights of the two samples of bipolar PNe.  
  }
\end{figure}

The larger physical size of R-BPNe with respect to W-BPNe
(Fig.~\ref{plot2}-{\it right}) can be interpreted as evidence
of the 
later evolutionary stage of R-BPNe with respect to W-BPNe.

To further confirm this observational fact, we have examined the PNe in the
sample listed by \citet{Frew2016} and selected those with W-BPN and R-BPN
morphologies.
Their sizes have been derived using the angular sizes listed in this
same reference and compared in Figure~\ref{bpn}-\emph{bottom}.  
This plot, based on a larger data sample, indeed confirms that
R-BPNe are larger than W-BPNe.

There is limited information on the kinematics of the PNe in Table~3.  
Kinematic studies are only available for two W-BPNe
\citep[M\,2-46 and NGC\,6881,][]{Manchado1996,Guerrero1998}, but for nine
R-BPNe \citep[Hen\,2-428, K\,3-58, K\,4-55, M\,1-59, M\,1-75, M\,2-52,
M\,4-14, Necklace, and Pr\'\i ncipe de Asturias,][]{Weinberger1989,Guerrero1996,Pena2002,Mampaso2006,Dobrincic2008,Santander2010,Corradi2011}.  
Using the new distances listed in Table~3, the innermost lobes of
W-BPNe have kinematic ages between 1300 yrs (NGC\,6881) and 6300
yrs (M\,2-46), whereas for R-BPNe these range from 2100 yrs
(Pr\'\i ncipe de Asturias) to 13800 yrs (K\,4-55), with a mean
kinematic age of 8000 yrs.
These figures also suggest that R-BPNe are older than W-BPNe.

The larger H$_2$ luminosity and H$_2$/Br$\gamma$ ratio of R-BPNe with
respect to W-BPNe can also be
interpreted into an evolutionary scheme.  
The dominant H$_2$ excitation mechanism in young PNe is UV fluorescence, 
because the density is not too high to collisionally de-excite the H$_2$ 
molecules as in proto PNe (PPNe), but the nebular size is not too large to
dilute the UV radiation field \citep{Natta1998}.  
Later on, as the PN keeps growing and the stellar luminosity declines, 
the local intensity of the UV radiation field is too low to produce 
significant UV excitation, whereas thermal emission from shock-excited 
hot H$_2$ molecules prevails \citep{Natta1998}.  
The collisional excitation in the ionized regions of 
PNe with high CSPN temperatures \citep{Aleman2011} 
can certainly enhance this excitation among evolved 
PNe.  
These trends are observationally confirmed in samples of PNe at
different evolutionary stages \citep{Davis2003}.

Can W-BPNe evolve into R-BPNe?
\citet{Balick1987} presented an evolutionary scheme for PNe, defining
\emph{``early''}, \emph{``middle''}, and \emph{``late''} PNe according
to the relative distance of the bright inner rim to the central star.  
In this scheme, late bipolar PNe develop large bipolar lobes and achieve large
aspect ratios.
The average aspect ratios of R-BPNe (0.70$\pm$0.18) and W-BPNe
(0.60$\pm$0.27) are, however, very similar.  
Actually, detailed hydrodynamic simulations of bipolar PNe do not
support this interpretation of the evolution of bipolar PNe, as they
achieve a shape early in their evolution that keeps growing with time
\citep{Icke1992} which is determined by the azimuthal density enhancement
of the AGB wind \citep{Frank1993}.  
In this paradigm, W-BPNe do not evolve into R-BPNe, 
although it has been recognised that hydrodynamic models have great difficulties to reproduce the most axisymmetric bipolar PNe \citep{Balick2000}.
Other shaping agents \citep[for instance, fast-moving knots and collimated outflows][]{Sahai1998,Dennis2008} have been proposed to be responsible of 
these extreme morphologies, but the time-evolution of the detailed morphology 
predicted in these scenarios has not been sufficiently explored.

Alternatively, W-BPNe and R-BPNe can evolve from different progenitor
populations.
This can be investigated by looking at the Galactic height distributions
of R-BPNe and W-BPNe shown in Figure~\ref{bpn}-\emph{top}.
R-BPNe and W-BPNe are mostly concentrated towards the Galactic Plane, with
mean Galactic heights of 290$\pm$310 pc and 300$\pm$270 pc, respectively.  
As first described by \citet{Corradi1995} and later on by \citet{Kastner1996},
bipolar H$_2$-emitting PNe are concentrated towards the Galactic Plane.  
Although the Galactic height distribution of W-BPNe have a secondary
peak at $\sim$700 pc, both samples have statistically consistent
Galactic height distribution, indicating that they can be drawn from
the same population of progenitor stars.

Even if W-BPNe and R-BPNe descend from the same progenitor population
(relatively massive low- and intermediate-mass stars), they may have
different progenitors.
\citet{Sahai1991} proposed a two-fold formation mechanism for bipolar PNe, 
arguing that equatorial tori results from 'born again' disks formed through
the destruction of planetary systems at the end of the AGB evolutionary
phase.  

The original amounts of dust in these sources can be very high, 
resulting in higher rates of H$_2$ formation, although 
R-BPNe show a conspicuous absence of dust thermal emission.  

Finally, it must be noted that a number of sources in our sample with low H$_2$/Br$\gamma$ ratio, 
namely PN\,G035.4$+$03.4, PN\,G038.9$−$01.3, PN\,G054.2$−$03.4, Hen\,2-428,
and M\,1-91, have questionable PN nature.
In this sense, it must be noted that the morphology of W-BPNe is
strikingly similar to that of sources classified as SS, such as
M\,2-9 \citep{Schmeja2001,Rodriguez2001,SmithGehrz2005,Clyne2015,Parker2016},
and that the scale height distribution of SS is similar to that
of bipolar PNe \citep{Boiarchuk1975,Bel2000}.  
PN\,G035.4$+$03.4 and M\,1-91 have been classified as SS, whereas the
PN nature of PN\,G038.9$−$01.3 has been questioned \citep{Parker2016}.
Otherwise, the central star of Hen\,2-428 has been claimed to harbor
a short-period double-degenerate core with combined mass above the
Chandrasekhar limit \citep{Santander2015}, whereas PN\,G054.2$−$03.4 
(the Necklace Nebula) is suspected to have gone through a common
envelope phase \citep{Miszalski2013}.

\section{Conclusion}

We have obtained deep near-IR observations of a sample of faint bipolar
PNe selected from the IPHAS sample to search for H$_2$ emission.
Molecular H$_2$ emission is found in most of them, mostly associated
with the brightest nebular regions in equatorial rings, but also in
bipolar lobes.

The present work also strengthens the dichotomy between H$_2$-bright
R-BPNe and H$_2$-weak W-BPNe.
A sample from the literature has been assembled to further investigate
these two sub-classes of bipolar PNe.
The excitation of W-BPNe is most likely caused by UV pumping, whereas
R-BPNe are mostly excited by shocks.  
W-BPNe also show continuum emission at their central regions,
but this is absent in R-BPNe.  
W-BPNe are not only intrinsically fainter in H$_2$ than R-BPNe, but they are
also smaller and have lower kinematic ages.
This suggests that W-BPNe are younger than R-BPNe, but it does not
necessarily implies that W-BPNe evolve into R-BPNe.
On the other hand, it seems they proceed from the same population of
progenitor stars.
The nature of many of those H$_2$-weak bipolar PNe has been disputed,
with a notable contamination of SS and post-common envelope binary
systems.

The H$_2$ emission from bipolar PNe with broad equatorial rings is not
associated with a classical PDR, but it rather comes from a discrete
distribution of compact knots and filaments.
The H$_2$ emission from these features is long-living, as the
H$_2$ molecules in them survive long in the nebular evolution.
As a result, their H$_2$ luminosity does not vary with the nebular expansion,
i.e., with the nebular age.

\section*{Acknowledgements}

GR-L acknowledge support from Universidad de Guadalajara
(Apoyo de Estancias Acad\'emicas -- RG/003/2017), CONACyT, CGCI, PRODEP and SEP (Mexico).
MAG acknowledges support of the grant AYA 2014-57280-P, co-funded with FEDER
funds.
He also acknowledges the hospitality of the Instituto de Astronom{\'i}a
y Meteorolog{\'i}a of the Universidad de Guadalajara (Jalisco, Mexico).  
LS acknowledges support from PAPIIT grant IA-101316 (Mexico).
ES acknowledge support from CONACyT and CUCEI (Mexico).
This research has made use of the HASH PN database at hashpn.space. 
This paper makes use of data obtained as part of the INT Photometric H$\alpha$ Survey of the Northern Galactic Plane (IPHAS) carried out at the Isaac Newton Telescope (INT). The INT is operated on the island of La Palma by the Isaac Newton Group in the Spanish Observatorio del Roque de los Muchachos of the Instituto de Astrofisica de Canarias. All IPHAS data are processed by the Cambridge Astronomical Survey Unit, at the Institute of Astronomy in Cambridge.
We thank Dr.\ Quentin Parker
for detailed reading of the manuscript and useful comments. 

\bibliographystyle{mn2e}

\begin{thebibliography}{}


\bibitem[\protect\citeauthoryear{Acosta Pulido et al.}{2003}]{Acosta2003} 
Acosta Pulido J.~A., et al., 2003, INGN, 7, 15 

\bibitem[\protect\citeauthoryear{Akras, Gon{\c c}alves, \& Ramos-Larios}{2017}]{Akras2017}
Akras S., Gon{\c c}alves D.~R., Ramos-Larios G., 2017, MNRAS, 465, 1289 

\bibitem[\protect\citeauthoryear{{Aleman} \& {Gruenwald}}{{Aleman} \&
  {Gruenwald}}{2011}]{Aleman2011}
{Aleman} I.,  {Gruenwald} R.,  2011, \aap, 528, A74

\bibitem[\protect\citeauthoryear{Balick}{1987}]{Balick1987}
  Balick B., 1987, AJ, 94, 671 

\bibitem[\protect\citeauthoryear{Balick}{2000}]{Balick2000}
Balick B., 2000, ASPC, 199, 41 

\bibitem[\protect\citeauthoryear{Barentsen et al.}{2014}]{Barentsen2014} 
Barentsen G., et al., 2014, MNRAS, 444, 3230

\bibitem[\protect\citeauthoryear{Beckwith, Gatley, \& Persson}{1978}]{Beckwith1978}
Beckwith S., Gatley I., Persson S.~E., 1978, ApJ, 219, L33 

\bibitem[\protect\citeauthoryear{Belczy{\'n}ski et al.}{2000}]{Bel2000}
Belczy{\'n}ski K., Miko{\l}ajewska J., Munari U., Ivison R.~J., Friedjung M.,
2000, A\&AS, 146, 407 

\bibitem[\protect\citeauthoryear{{Black} \& {van Dishoeck}}{{Black} \& {van
  Dishoeck}}{1987}]{Black1987}
{Black} J.~H.,  {van Dishoeck} E.~F.,  1987, \apj, 322, 412

\bibitem[\protect\citeauthoryear{Bohlin, Savage, \& Drake}{1978}]{Bohlin1978}
Bohlin R.~C., Savage B.~D., Drake J.~F., 1978, ApJ, 224, 132 

\bibitem[\protect\citeauthoryear{Boiarchuk}{1975}]{Boiarchuk1975}
Boiarchuk A.~A., 1975, IAUS, 67, 377 

\bibitem[\protect\citeauthoryear{{Burton}, {Hollenbach} \& {Tielens}}{{Burton}
  et~al.}{1992}]{Burton1992}
{Burton} M.~G.,  {Hollenbach} D.~J.,    {Tielens} A.~G.~G.,  1992, \apj, 399,
  563

\bibitem[\protect\citeauthoryear{Clyne et al.}{2015}]{Clyne2015}
Clyne N., Akras S., Steffen W., Redman M.~P., Gon{\c c}alves D.~R.,
Harvey E., 2015, A\&A, 582, A60 

\bibitem[\protect\citeauthoryear{{Corradi}, {Sabin}, {Miszalski},
  {Rodr{\'{\i}}guez-Gil}, {Santander-Garc{\'{\i}}a}, {Jones}, {Drew}, {Mampaso}
  \& {et al.}}{{Corradi} et~al.}{2011}]{Corradi2011}
{Corradi} R.~L.~M.,  {Sabin} L.,  {Miszalski} B.,  {Rodr{\'{\i}}guez-Gil} P.,
  {Santander-Garc{\'{\i}}a} M.,  {Jones} D.,  {Drew} J.~E.,  {Mampaso} A.,
  {et al.} 2011, \mnras, 410, 1349

\bibitem[\protect\citeauthoryear{{Corradi} \& {Schwarz}}{{Corradi} \&
  {Schwarz}}{1995}]{Corradi1995}
{Corradi} R.~L.~M.,  {Schwarz} H.~E.,  1995, \aap, 293, 871

\bibitem[\protect\citeauthoryear{Cox et al.}{Cox et al.}{1998}]{Cox1998}
  Cox, P.\ et al.\ 1998, ApJ, 495, L23

\bibitem[\protect\citeauthoryear{Davis et al.}{2003}]{Davis2003}
  Davis C.~J., Smith M.~D., Stern L., Kerr T.~H., Chiar J.~E., 2003,
  MNRAS, 344, 262 
	
\bibitem[Dennis et al.(2008)]{Dennis2008} 
Dennis, T.~J., Cunningham, A.~J., Frank, A., et al.\ 2008, \apj, 679, 1327-1337 

\bibitem[\protect\citeauthoryear{Dickey \& Lockman}{1990}]{Dickey1990}
Dickey J.~M., Lockman F.~J., 1990, ARA\&A, 28, 215 

\bibitem[\protect\citeauthoryear{{Dinerstein}, {Lester}, {Carr} \&
  {Harvey}}{{Dinerstein} et~al.}{1988}]{Dinerstein1988}
{Dinerstein} H.~L.,  {Lester} D.~F.,  {Carr} J.~S.,    {Harvey} P.~M.,  1988,
  \apjl, 327, L27

\bibitem[\protect\citeauthoryear{Dobrin{\v c}i{\'c} et al.}{2008}]{Dobrincic2008}
  Dobrin{\v c}i{\'c} M., Villaver E., Guerrero M.~A., Manchado A., 2008, AJ, 135, 2199 

\bibitem[\protect\citeauthoryear{Drew et al.}{2005}]{Drew2005} 
Drew J.~E., et al., 2005, MNRAS, 362, 753

\bibitem[\protect\citeauthoryear{Frank et al.}{1993}]{Frank1993}
Frank A., Balick B., Icke V., Mellema G., 1993, ApJ, 404, L25 

\bibitem[\protect\citeauthoryear{Frew, Parker, \& Boji{\v c}i{\'c}}{2016}]{Frew2016}
Frew D.~J., Parker Q.~A., Boji{\v c}i{\'c} I.~S., 2016, MNRAS, 455, 1459 

\bibitem[\protect\citeauthoryear{Froebrich et al.}{2015}]{Froebrich2015}
Froebrich D., et al., 2015, MNRAS, 454, 2586   

\bibitem[\protect\citeauthoryear{Gledhill et al.}{2017}]{Gledhill2017}
Gledhill, T.M., et al.\ 2017, IAU Symp.\ 323, Planetary Nebulae: Multi-Wavelength Probes of Stellar and Galactic Evolution, \textsl{in press}.

\bibitem[\protect\citeauthoryear{Gonz{\'a}lez-Solares et al.}{2008}]{Gonzalez2008} 
Gonz{\'a}lez-Solares E.~A., et al., 2008, MNRAS, 388, 89
  
\bibitem[\protect\citeauthoryear{Guerrero \& Manchado}{1998}]{Guerrero1998}
   Guerrero M.~A., Manchado A., 1998, ApJ, 508, 262 

\bibitem[\protect\citeauthoryear{Guerrero, Manchado, \& Serra-Ricart}{1996}]{Guerrero1996}
   Guerrero M.~A., Manchado A., Serra-Ricart M., 1996, ApJ, 456, 651 

\bibitem[\protect\citeauthoryear{{Guerrero}, {Villaver}, {Manchado},
  {Garcia-Lario} \& {Prada}}{{Guerrero} et~al.}{2000}]{Guerrero2000}
{Guerrero} M.~A.,  {Villaver} E.,  {Manchado} A.,  {Garcia-Lario} P.,
  {Prada} F.,  2000, \apjs, 127, 125

\bibitem[\protect\citeauthoryear{{Hora} \& {Latter}}{{Hora} \&
  {Latter}}{1994}]{Hora1994}
{Hora} J.~L.,  {Latter} W.~B.,  1994, \apj, 437, 281

\bibitem[\protect\citeauthoryear{Hora \& Latter}{1996}]{Hora1996}
  Hora J.~L., Latter W.~B., 1996, ApJ, 461, 288 

\bibitem[\protect\citeauthoryear{Hora, Latter, \& Deutsch}{1999}]{Hora1999}
  Hora J.~L., Latter W.~B., Deutsch L.~K., 1999, ApJS, 124, 195 

\bibitem[\protect\citeauthoryear{{Huggins}, {Bachiller}, {Cox} \&
  {Forveille}}{{Huggins} et~al.}{1996}]{Huggins1996}
{Huggins} P.~J.,  {Bachiller} R.,  {Cox} P.,    {Forveille} T.,  1996, \aap,
  315, 284

\bibitem[\protect\citeauthoryear{Icke, Balick, \& Frank}{1992}]{Icke1992}
Icke V., Balick B., Frank A., 1992, A\&A, 253, 224

\bibitem[\protect\citeauthoryear{Kalberla et al.}{2005}]{Kalberla2005}
Kalberla P.~M.~W., Burton W.~B., Hartmann D., Arnal E.~M., Bajaja E., Morras R., P{\"o}ppel W.~G.~L., 2005, A\&A, 440, 775 

\bibitem[\protect\citeauthoryear{Karakas et al.}{2009}]{Karakas2009} 
Karakas A.~I., van Raai M.~A., Lugaro M., Sterling N.~C., Dinerstein H.~L., 2009, ApJ, 690, 1130 

\bibitem[\protect\citeauthoryear{Kastner et al.}{1994}]{Kastner1994} 
Kastner J.~H., Gatley I., Merrill K.~M., Probst R., Weintraub D., 1994, ApJ, 421, 600 

\bibitem[\protect\citeauthoryear{Kastner et al.}{1996}]{Kastner1996} 
Kastner J.~H., Weintraub D.~A., Gatley I., Merrill K.~M., Probst R.~G., 1996, ApJ, 462, 777

\bibitem[\protect\citeauthoryear{Likkel et al.}{2006}]{Likkel2006}
  Likkel L., Dinerstein H.~L., Lester D.~F., Kindt A., Bartig K., 2006,
  AJ, 131, 1515 

\bibitem[\protect\citeauthoryear{Mampaso et al.}{2006}]{Mampaso2006} 
Mampaso A., et al., 2006, A\&A, 458, 203 

\bibitem[\protect\citeauthoryear{Manchado et al.}{2003}]{Manchado2003} 
Manchado A., et al., 2003, RMxAC, 16, 43

\bibitem[\protect\citeauthoryear{Manchado, Stanghellini, \& Guerrero}{1996}]{Manchado1996}
  Manchado A., Stanghellini L., Guerrero M.~A., 1996, ApJ, 466, L95 
  
\bibitem[\protect\citeauthoryear{Manchado et al.}{2015}]{Manchado2015}
  Manchado A., Stanghellini L., Villaver E., Garc{\'{\i}}a-Segura G., Shaw R.~A.,
  Garc{\'{\i}}a-Hern{\'a}ndez D.~A., 2015, ApJ, 808, 115 

\bibitem[\protect\citeauthoryear{Marquez-Lugo et al.}{2015}]{Marquez2015}
  Marquez-Lugo R.~A., Guerrero M.~A., Ramos-Larios G., Miranda L.~F., 2015,
  MNRAS, 453, 1888 

\bibitem[\protect\citeauthoryear{{Marquez-Lugo}, {Ramos-Larios}, {Guerrero} \&
  {V{\'a}zquez}}{{Marquez-Lugo} et~al.}{2013}]{Marquez2013}
{Marquez-Lugo} R.~A.,  {Ramos-Larios} G.,  {Guerrero} M.~A.,    {V{\'a}zquez}
  R.,  2013, \mnras, 429, 973

\bibitem[\protect\citeauthoryear{{Matsuura}, {Zijlstra}, {Gray}, {Molster} \&
  {Waters}}{{Matsuura} et~al.}{2005}]{Matsuura2005}
{Matsuura} M.,  {Zijlstra} A.~A.,  {Gray} M.~D.,  {Molster} F.~J.,    {Waters}
  L.~B.~F.~M.,  2005, \mnras, 363, 628

\bibitem[\protect\citeauthoryear{Matsuura et al.}{2009}]{Matsuura2009}
Matsuura M., et al., 2009, ApJ, 700, 1067

\bibitem[\protect\citeauthoryear{{Miszalski}, {Boffin} \&
  {Corradi}}{{Miszalski} et~al.}{2013}]{Miszalski2013}
{Miszalski} B., {Boffin} H.~M.~J., {Corradi} R.~L.~M.,  2013, \mnras, 428, L39

\bibitem[\protect\citeauthoryear{{Natta} \& {Hollenbach}}{{Natta} \& {Hollenbach}}{1998}]{Natta1998}
{Natta} A.,  {Hollenbach} D.,  1998, \aap, 337, 517

\bibitem[\protect\citeauthoryear{Osterbrock \& Ferland}{2006}]{Osterbrock2006}
Osterbrock D.~E., Ferland G.~J., 2006, Astrophysics of Gaseous Nebulae and
Active Galactic Nuclei, University Science Books, Sausalito, California

\bibitem[\protect\citeauthoryear{Parker et al.}{2006}]{Parker2006} 
Parker Q.~A., et al., 2006, MNRAS, 373, 79 

\bibitem[\protect\citeauthoryear{Parker, Boji{\v c}i{\'c}, \& Frew}{2016}]{Parker2016}
Parker Q.~A., Boji{\v c}i{\'c} I.~S., Frew D.~J., 2016, JPhCS, 728, 032008 

\bibitem[\protect\citeauthoryear{{Peimbert} \& {Torres-Peimbert}}{{Peimbert} \&
  {Torres-Peimbert}}{1983}]{Peimbert1983}
{Peimbert} M.,  {Torres-Peimbert} S.,  1983, in {Flower} D.~R.,  ed., Planetary
  Nebulae Vol.~103 of IAU Symposium, {Type I planetary nebulae}.
pp 233--241

\bibitem[\protect\citeauthoryear{Pe{\~n}a \& Medina}{2002}]{Pena2002}
  Pe{\~n}a M., Medina S., 2002, RMxAA, 38, 23 

\bibitem[\protect\citeauthoryear{{Peretto}, {Fuller}, {Zijlstra} \&
  {Patel}}{{Peretto} et~al.}{2007}]{Peretto2007}
{Peretto} N.,  {Fuller} G.,  {Zijlstra} A.,    {Patel} N.,  2007, \aap, 473,
  207 

\bibitem[\protect\citeauthoryear{{Phillips} \& {Marquez-Lugo}}{{Phillips} \&
  {Marquez-Lugo}}{2011}]{Phillips2011}
{Phillips} J.~P.,  {Marquez-Lugo} R.~A.,  2011, \mnras, 410, 2257

\bibitem[\protect\citeauthoryear{{Phillips}, {White} \& {Harten}}{{Phillips}
  et~al.}{1985}]{Phillips1985}
{Phillips} J.~P.,  {White} G.~J.,    {Harten} R.,  1985, \aap, 145, 118

\bibitem[\protect\citeauthoryear{Ramos-Larios, Guerrero, \& Miranda}{2008}]{Ramos2008}
Ramos-Larios G., Guerrero M.~A., Miranda L.~F., 2008, AJ, 135, 1441 

\bibitem[\protect\citeauthoryear{{Ramos-Larios}, {Guerrero}, {V{\'a}zquez} \&
  {Phillips}}{{Ramos-Larios} et~al.}{2012}]{Ramos2012}
{Ramos-Larios} G.,  {Guerrero} M.~A.,  {V{\'a}zquez} R.,    {Phillips} J.~P.,
2012, \mnras, 420, 1977

\bibitem[\protect\citeauthoryear{Rodr{\'{\i}}guez, Corradi, \& Mampaso}{2001}]{Rodriguez2001}
Rodr{\'{\i}}guez M., Corradi R.~L.~M., Mampaso A., 2001, A\&A, 377, 1042 

\bibitem[\protect\citeauthoryear{{Sabin}}{{Sabin}}{2008}]{Sabin2008}
{Sabin} L.,  2008, PhD thesis, Jodrell Bank Centre for Astrophysics, University
of Manchester, UK

\bibitem[\protect\citeauthoryear{Sabin et al.}{2014}]{Sabin2014} 
Sabin L., et al., 2014, MNRAS, 443, 3388

\bibitem[\protect\citeauthoryear{Sahai \& Trauger}{1998}]{Sahai1998}
Sahai R., Trauger J.~T., 1998, AJ, 116, 1357 

\bibitem[\protect\citeauthoryear{Sahai et al.}{1991}]{Sahai1991}
Sahai R., Wootten A., Schwarz H.~E., Clegg R.~E.~S., 1991, A\&A, 251, 560 

\bibitem[\protect\citeauthoryear{Santander-Garc{\'{\i}}a et al.}{2010}]{Santander2010}
  Santander-Garc{\'{\i}}a M., et al., 2010, A\&A, 519, A54 

\bibitem[\protect\citeauthoryear{Santander-Garc{\'{\i}}a et al.}{2015}]{Santander2015}
Santander-Garc{\'{\i}}a M., Rodr{\'{\i}}guez-Gil P., Corradi R.~L.~M., Jones D., Miszalski B., Boffin H.~M.~J., Rubio-D{\'{\i}}ez M.~M., Kotze M.~M., 2015, Natur, 519, 63 

\bibitem[\protect\citeauthoryear{Schmeja \& Kimeswenger}{2001}]{Schmeja2001}
Schmeja S., Kimeswenger S., 2001, A\&A, 377, L18 

\bibitem[\protect\citeauthoryear{{Smith}, {Balick} \& {Gehrz}}{{Smith}
  et~al.}{2005}]{Smith2005}
{Smith} N.,  {Balick} B.,    {Gehrz} R.~D.,  2005, \aj, 130, 853

\bibitem[\protect\citeauthoryear{Smith \& Gehrz}{2005}]{SmithGehrz2005}
Smith N., Gehrz R.~D., 2005, AJ, 129, 969 

\bibitem[\protect\citeauthoryear{Speck et al.}{2002}]{Speck2002} 
Speck A.~K., Meixner M., Fong D., McCullough P.~R., Moser D.~E., Ueta T., 2002, AJ, 123, 346 

\bibitem[\protect\citeauthoryear{{Sternberg} \& {Dalgarno}}{{Sternberg} \&
  {Dalgarno}}{1989}]{Sternberg1989}
{Sternberg} A.,  {Dalgarno} A.,  1989, \apj, 338, 197

\bibitem[\protect\citeauthoryear{Viironen et al.}{2009}]{Viironen2009} 
Viironen K., et al., 2009, A\&A, 504, 291

\bibitem[\protect\citeauthoryear{{Webster}, {Payne}, {Storey} \&
  {Dopita}}{{Webster} et~al.}{1988}]{Webster1988}
{Webster} B.~L.,  {Payne} P.~W.,  {Storey} J.~W.~V.,    {Dopita} M.~A.,  1988,
  \mnras, 235, 533

\bibitem[\protect\citeauthoryear{Weinberger}{1989}]{Weinberger1989}
  Weinberger R., 1989, A\&AS, 78, 301 

\bibitem[\protect\citeauthoryear{{Young}, {Cox}, {Huggins}, {Forveille} \&
  {Bachiller}}{{Young} et~al.}{1999}]{Young1999}
{Young} K.,  {Cox} P.,  {Huggins} P.~J.,  {Forveille} T.,    {Bachiller} R.,
1999, \apj, 522, 387

\bibitem[\protect\citeauthoryear{Zuckerman \& Gatley}{1988}]{Zuckerman1988}
Zuckerman B., Gatley I., 1988, ApJ, 324, 501 

\end{thebibliography}

\label{lastpage}

\end{document}